\documentclass[pre,aps,amsmath,eqsecnum,nofootinbib]{revtex4-1}
\usepackage{amssymb}
\usepackage{amsmath} 
\usepackage{graphicx}
\usepackage{amsbsy}
\usepackage{color}
\usepackage{bm}
\usepackage{soul}

\begin{document}
\title{Theory and simulation studies of self-assembly of helical particles}
\author{Giorgio Cinacchi} 
\email{giorgio.cinacchi@uam.es}
\affiliation{
Dep artamento de F\'{i}sica Te\'{o}rica de la Materia Condensada, 
Instituto de F\'{i}sica de la Materia Condensada (IFIMAC) and
Instituto de Ciencias de Materiales "Nicol\'{a}s Cabreras", 
Universidad Aut\'{o}noma de Madrid,
Campus de Cantoblanco, 28049 Madrid, Spain
}
\author{Alberta Ferrarini} 
\email{alberta.ferrarini@unipd.it}
\affiliation{Dipartimento di Scienze Chimiche, 
Universit\`{a} di Padova, via F. Marzolo 1, 35131 Padova, Italy}

\author{Elisa Frezza}
\affiliation{BMSSI, UMR 5086 CNRS/Univ. Lyon I, Institut de Biologie et Chimie des Prot\'eines,
7 passage de Vercours, Lyon 69367, France}

\author{Achille Giacometti} 
\email{achille.giacometti@unive.it}
\affiliation{Dipartimento di Scienze Molecolari e Nanosistemi, 
Universit\`{a} Ca' Foscari di Venezia,
Campus Scientifico, Edificio Alfa,
via Torino 155, 30170 Venezia Mestre, Italy}

\author{Hima Bindu Kolli} 
\affiliation{Department of Chemistry, University of Oslo,
Postboks 1033 Blindern 0315 Oslo, Norway}

\date{\today}

\begin{abstract}

\end{abstract}
\maketitle


\tableofcontents 



\section{Introduction: why hard helices?}
\label{sec:introduction}
\noindent Crystallization as an ordering phenomenon is neither restricted to molecular systems, nor to attractive interactions \cite{Whitesides02}. Even in the absence of explicit attractive interactions, systems might be spontaneously forming ordered aggregates driven by entropic gain, a well known example being the formation of a crystal structure in a hard colloidal sphere system \cite{Pusey86}. 
This is one of the several counterintuitive examples where there is a spontaneous ordering driven by entropy  \cite{Frenkel14}. While in this case the number of possible
crystal structures was still limited by the aspecificity of the steric interactions, the newly developments in the chemical synthesis of colloids paved the way
to construct crystal with prescribed crystal structures by changing the shape of the colloidal bulding blocks \cite{Glotzer07}.

Together with the shape, also the chirality of the building blocks may play an important, and yet not fully explored, role in self-assembly processes \cite{Damasceno15}. Leaving aside the issue of
the emergence of homochirality in biological systems from equally probable molecular chiral  moieties, a crucial problem in the current understanding
of the origin of life, the use of chiral particles finds important applications in photonic metamaterials.
Helically nanostructured materials have been attracting increasing attention, also because of their unique electrical and mechanical properties \cite{Yang,Fisher,Liu14}.
The control in their three-dimensional organization is also a crucial step in pushing the chiral properties to a mesoscale range.  

While it is possible to control the enantioselective process by using depletion interactions and faceting of the bulding blocks \cite{Damasceno15}, helices are among the natural objects to focus on when dealing with chirality.
New functional materials \cite{Douglas09} \cite{Seeman03} can be produced by exploiting the intrinsic chirality of the helical structures, which are useful in  catalysis and demixing of enantiomers 
\cite{Nakano01} \cite{Yashima09}.  The importance of the helix in nature is unquestionable: proteins, polysaccharides, DNA and RNA, the so called molecules of life, have a helical structure.
The helical shape is exhibited in nature also by microorganisms, like filamentous viruses, and cell organelles, like bacterial flagella. 
Filamentous viruses  are formed by a DNA of RNA core, wrapped by a coating of helically arranged proteins.
Well-known examples are Tobacco Mosaic Virus (TMV), the first discovered virus \cite{tmv}, 
 and viruses related to filamentous phage \emph{fd}, whose mutants are present in nature (M13, fd), while others can be  
 obtained by genetic engineering. They have been widely investigated as models of highly anisotropic,  colloidal systems, with the advantage of being essentially monodisperse and that their length, of the order of a micrometer, makes them 
suitable for imaging techniques, such as optical microscopy. 
Bacterial flagella are helical macromolecular structures assembled from a single protein (flagellin). 
Their helical shape can be tuned with high precision
by regulating external parameters such as temperature or pH, and  their large size, of the order of microns, makes them very handy for optical observations.

Because of their shape anisotropy, helical biopolymers and colloidal particles may exhibit liquid crystal phases at high densities \cite{Rey10}.  These phases are often tacitly assumed to be the same as 
those occurring in systems of rod-like particles. 
However, it cannot be taken for granted that at such high densities the intrinsic helicity of the shape can be neglected.
To explore the effect of self-assembly of helical polymers and colloids, and in particular to discover whether there is anything special just determined by the helical shape,  
we have undertaken a comprehensive investigation of the phase behavior of hard helices\cite{Frezza13,Kolli14a,Frezza14,Kolli14b,Kolli15}, 
interacting through purely steric repulsions, using  Monte Carlo simulations  and  an extension of Onsager theory 
\cite{Onsager}, a density functional theory (DFT) that was originally proposed to explain the onset if nematic 
ordering in a system of hard rods.
These studies have revealed an unexpectedly rich phase behavior,
the most interesting result being the existence of special phases characterized by \emph{screw}-like ordering. 
Such kind of organization had been proposed for DNA, based on theoretical considerations \cite{Manna07}, 
and had been observed in dense suspensions of flagellar filaments \cite{Barry06}. 
Hard helices are an athermal system: phase transitions are controlled by density and 
are driven by the entropy gain on moving from the less to the more ordered phase. 
This is a minimalist model, possibly insufficient to account entirely for the complexity of real systems. 
It is nonetheless useful to obtain a general picture, lacking in the previous literature, and represents a useful reference for interpreting the behavior of systems that may be  more complicated.  

The remaining of this chapter  is organized as follows. In Sec. \ref{sec:LC} 
we briefly recall the main features of liquid crystal phases, with special attention to their chiral versions.
Then, we will describe our model systems and the methods used to investigate them. In Sec. \ref{sec:onsager} we give an outline of Onsager theory, whereas in Sec. \ref{sec:DFThelical} we extend Onsager theory to helically modulated nematic phases. In Sec. \ref{sec:OP} we review the definitions of the order parameters and correlation functions necessary to identify the various phases. 
In Sec. \ref{sec:origin} we discuss the physical origin of the chiral nematic phases formed by helical particles, 
and the next Section (\ref{sec:phase}) gives an overview of the phase behavior of hard helices. 
In Sec. \ref{sec:exp} theoretical predictions are compared to experimental data and 
finally Sec. \ref{sec:conclusions} presents the conclusions and possible perspectives.

\section{Liquid crystal phases}
\label{sec:LC}
\noindent  Anisometric molecules and particles can form liquid crystal phases, which are fluid states characterized by long-range orientational order, while  long-range translational order is absent or only partial. 
The phase transitions can be induced by changes of temperature or density, and in the two cases one speaks of  thermotropic and lyotropic systems, respectively. To the former class, mostly represented by organic low molar mass molecules and polymers, belong the materials used in electro-optical applications. The latter class includes various kinds of systems, like surfactants, lipids, anisotropic colloids and stiff  or semi-flexible polymers. Important examples of lyotropic liquid crystals can be found in nature, e.g. biomembranes and DNA.     

Liquid crystals comprise a variety of phases, differing from each other in the kind and symmetry of order \cite{deGennesbook,HandbookLC}.  Probably the 
most common is the \emph{nematic} (N), in which  the centers of mass of molecules or particles are randomly distributed in space, but their long axes are preferentially aligned to each other. The average alignment axis is denoted as the \emph{director} ($\widehat{\mathbf{n}}$).  This is the phase 
generally found 
in the proximity of the isotropic phase, 
at lower  temperature in thermotropic or
higher density in lyotropic systems. 
With further decreasing temperature or increasing density, smectic (Sm) phases may be found, which exhibit and additional one-dimensional order: 
molecules or particles are preferentially located in layers, 
and the director may be either parallel (e.g smectic A, smectic B) or tilted (e.g. smectic C) with respect to the layer normal ($Z$ axis).
In the smectic A and C phases the positions of molecules or particles are randomly distributed within layers,
whereas in the smectic B phase there is hexatic short-range order within the layers.
Other, less common phases can be found for specific systems. 
One of them is the \emph{biaxial} N phase (N$_b$), which differs form the conventional ('uniaxial') N phase because the orientational distribution in the plane perpendicular to the director  $\widehat{\mathbf{n}}$ is  anisotropic: thus two other directors,   
$\widehat{\mathbf{b}}$ and $\widehat{\mathbf{c}}$ can be defined, with 
$\widehat{\mathbf{n}}=\widehat{\mathbf{c}}\times \widehat{\mathbf{b}}$ \cite{biaxialbook,Zannoni}.
Biaxial  N phases (N$_b$) were detected in solutions of surfactants that self-assemble into biaxial micelles  \cite{Yu} and in colloidal suspensions of board-like particles \cite{Vroege}, whereas their existence in thermotropic systems is more controversial.
In this context, bent mesogens were proposed as  suitable candidates, but the results are less straightforward. 
 
Chiral molecules or particles can impart  the liquid crystal phase a chiral character  \cite{Chiralitybook}. 
The chiral analog of the nematic is 
the \emph{cholesteric} or \emph{twisted nematic} (N$^\ast$) phase, where the director
 $\widehat{\mathbf{n}}$ rotates in helical way around a perpendicular axis, 
rather than being uniform. 
Handedness and pitch ($\cal P$) of the cholesteric helix are determined by the 
structure at the molecular level, 
but the connection is not straightforward  \cite{Spada11,Katsonis}. The general features are that pitches are orders of magnitude longer than the molecular size (from hundreds of nanometers to millimeters) and 
cholesteric phases formed by enantiomers have identical pitch and opposite handedness. 
Some chiral systems exhibit also, between the isotropic and the cholesteric phase, one or more Blue Phases \cite{Wright89}, which can be described as fluid lattices of defects, with cubic symmetry and   
lattice periods of the order of the wavelength of visible light. 
They are locally chiral, since directors are locally arranged in double-twist cylinders. 
On the other boundary of the N$^\ast$ phase, also the smectic phases may be chiral: in the so-called smectic C chiral
 (SmC$^\ast$) phase the director, tilted with respect to the layer normal, rotates in helical way from layer to layer,
 again with typical pitches longer than 100 nm. In the case of short-pitch cholesterics, between the cholesteric and
 the smectic, \emph{twist grain boundary} (TGB) phases may be found. These are frustrated structures, 
first predicted by de Gennes \cite{deGennes} and Lubensky \cite{Lubensky}, 
and observed soon after \cite{Goodby},  composed of smectic slabs, 
rotated with respect to one another and separated by defect walls.
Such a helical superstructure results from the competition between the cholesteric organization  and 
smectic layering, 
which cannot be simultaneously realized without the formation of defects. 

\section{Hard helices: a minimal model }
\label{sec:minimal}
\begin{figure}
\centering
\includegraphics[width=5cm]{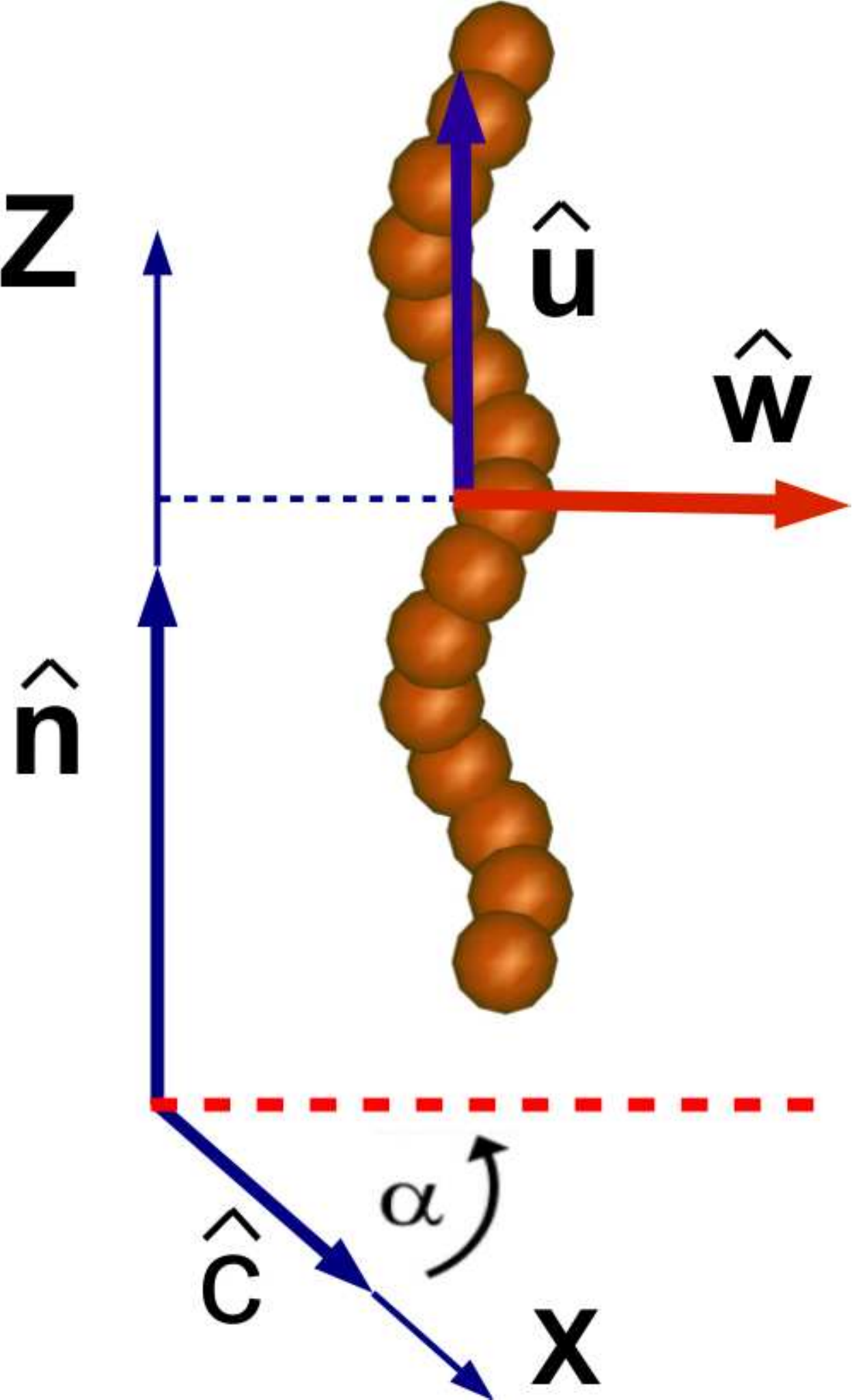}
\caption{Model helix made up of a chain of partially overlapping  hard spheres.
The orientation of the helix in space is univocally defined by the direction of its 
main axis $\widehat{\mathbf{u}}$ and its two-fold symmetry axis
$\widehat{\mathbf{w}}$. 
Reproduced from Ref. \cite{Kolli14b} with permission from the Royal Society of Chemistry.
}
\label{fig:helix}
\end{figure}
\begin{figure}	
\centering
\includegraphics[width=10cm]{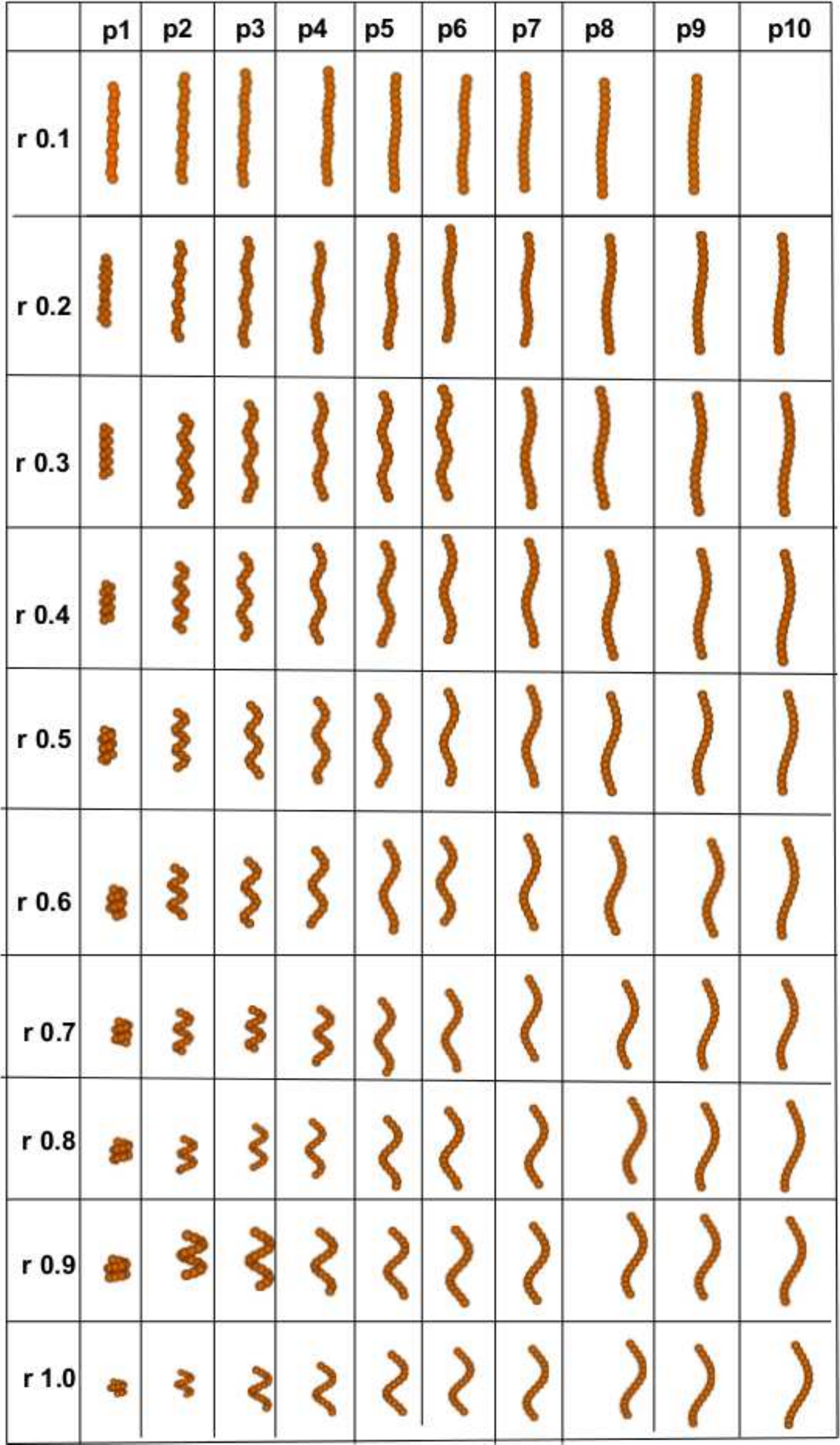}
\caption{Helical shapes studied, with radius $r$ ranging from 0.1 to 1.0, pitch $p$ ranging from 1 to 10, and constant contour length $L=10$.}
\label{fig:helixshapes}
\end{figure}

\noindent Our model helices are formed by chains of $N_{b}$ partially fused hard spherical beads, each of diameter $D$, rigidly arranged in a right-handed helical fashion as shown in Figure \ref{fig:helix}. 
All lengths will eventually be expressed in units of $D$ that will set the length scale. The morphology of the helix will be defined by providing the radius $r$ and the pitch $p$, along with
an additional parameter that can be chosen either to be the contour length $L$, or the euclidean length $\Lambda$. Upon fixing the contour length $L$ and changing $r$ and $p$ independently, 
the aspect ratio will be changing. 
Conversely, upon fixing  the euclidean length $\Lambda$ and changing $r$ and $p$ independently, the contour length $L$ will be modified. 
We can easily find the relation between $\Lambda$ and $L$. The centres of the beads can be identified by the following helix parametric equations 
\begin{align}
\label{sec2:eq1}
x_i =& r \cos(2 \pi t_i)\nonumber \\ 
y_i =& r \sin(2 \pi t_i),\qquad 1 \le i \le N_{b} \\ 
z_i =& p t_i \nonumber
\end{align}
the centres of the beads lie on an inner cylinder of radius $r$, whereas the diameter of the outer cylinder (2$r$ + $D$) is the width of the helix $r_{max}$. 
The long axis of the helix $\hat{\mathbf{u}}$ passes through the center of the helix.
Given $r$, $p$ and $L$, the increment $\Delta t = t_{i+1} - t_i$ can be computed as
\begin{equation}
\frac{L}{14}=2 \pi \Delta t \sqrt{r^2+ \left(\frac{p}{2 \pi}\right)^2}
\end{equation}
The parameteric equations could alternatively be written using number of turns as the fixed value instead of fixed $L$ \cite{Frezza13}. The euclidean length $\Lambda$ is measured as the component parallel to the long axis of the helix of the distance between first and last bead. 
Different helix shapes  -- from a slender rod to a highly coiled helix -- can be achieved upon varying $r$ and $p$ independently, as illustrated in Figure \ref{fig:helixshapes}.
The limit case of $r=0$ corresponds to the a
 rod-like shape that can be contrasted with known results from the phase diagram of hard spherocylinders \cite{Bolhuis97}.  

In our studies we have generally focused on helices of $N_b=15$ beads and constant contour length $L=10$ (see Figure \ref{fig:helixshapes}).

 In a system of our model helices, the only interactions are hard-core repulsions, that is beads belonging to different helices interact as follows 
\begin{align}
u(r_{ij}) =&  \infty,  \quad r_{ij} < D \\
u(r_{ij}) =&  0,        \quad r_{ij} > D
\end{align}
where $r_{ij}$ is the the distance between a pair of beads belonging to different helices.

\section{Numerical simulations}
\label{sec:numerical}
\noindent The problem of calculating the properties of a condensed matter system by solving the equations of motion appears as a dreadful task in view of the large number of particles ($\approx 10^{23}$) involved.
Statistical mechanics helps to relate the equilibrium thermodynamic properties of the system to the average of particle properties by treating them in a probabilistic way, thus bridging microscopic properties
with thermodynamical quantities. In this framework, Monte Carlo (MC) methods have been evolved in the past decades as one of the most efficient and reliable tool for assessing, 
and calculating thermodynamical properties and drawing the corresponding phase diagrams. 

One of the crucial issues for generating  proper microstates is ergodicity. Macroscopic properties measured in experiments are time averaged quantities. The time scale of the observation 
is very large when compared to the switching time scale from one microstate to another. Because of this, we can assume that system visits all accessible microstates during the observation time. 
Hence a macroscopic quantity obtained by taking an average over a suitable ensemble  is equivalent to the time averaged quantity. In order to ensure that the system is in equilibrium, 
the sampling of microstates should obey ``detailed balance'', which requires the forward and backward rates of any transition between states to be equal.
   
Different ensembles, corresponding to the different independent thermodynamical variables can be used by performing appropriate Legendre transformations.
In our studies we have employed the Monte Carlo technique under isothermal-isochoric ($NVT$) and isothermal-isobaric conditions ($NPT$) to study the phase behaviour of systems of hard helices. 
Here $N$, $V$, $P$ and $T$ denote the number of particles, the total volume, the pressure, and the temperature of the system, respectively.

\subsection{Monte Carlo in various ensembles}
\noindent The first MC simulation was implemented on hard disks by Metropolis in 1953 \cite{Metropolis}. Since then MC has evolved into a reliable methodology for studying 
virtually all soft matter systems, ranging from simple and molecular fluids, to polymers, surfactants, and proteins. This list includes mesogens forming liquid crystal phases.

In general, a MC  approach consists of the three steps. (i)  Determining the microstate probability distribution for the ensemble of interest. (ii) Determining the set of MC moves accomplishing changes in all fluctuating quantities. (iii) Imposing the detailed balance condition to find the acceptance criterion.

\subsubsection{Canonical Monte Carlo simulations (NVT--MC)}
\noindent The most common and convenient ensemble for fluid systems is usually the canonical (NVT) ensemble, where $N$ particles are inserted into a fixed computational box of volume $V$ at a given temperature $T$. 
In this case. the pressure is computed via the virial theorem \cite{AllenTildesley}
\begin{eqnarray}
P &=& \rho k_B T + \frac{1}{3V} \left\langle \sum_{i<j} \mathbf{f}_{ij} \cdot (\mathbf{R}_i-\mathbf{R}_j)\right\rangle
\end{eqnarray}  
where as $\rho$ is the number density, and $\mathbf{f}_{ij}$ is the force between particles $i$ and $j$, located at  $\mathbf{R}_{i}$ and $\mathbf{R}_{j}$, respectively. 
For isotropic potentials, depending only on the distance $R_{ij}=\vert \mathbf{R}_{i}-\mathbf{R}_{j}\vert$ between particles,
this can be easily translated in terms of an integral involving the pair correlation function  $g(R)$. However, the same procedure is not as straightforward
for non-spherical hard-core discontinuous potentials. 

\subsubsection{Isothermal-isobaric Monte Carlo simulations (NPT--MC)}
\noindent Isothermal-isobaric (NPT) Monte Carlo simulations were first employed by W. W. Wood \cite{Wood} 
in the study of a fluid of hard disks. Isothermal-isobaric ensemble is widely used because most of the experiments are done at controlled pressure and temperature. In NPT--MC simulations the system is assumed to be 
in thermal contact with a large heat bath, and mechanically coupled with a barostat. The mechanical coupling allows the system to change its volume in order to keep its pressure constant.
The problem of finding the correct probability distribution (Point (i) above) is a standard textbook topic \cite{AllenTildesley}, so we will here focus on the problem of finding the correct moves for generic 
biaxial particles. 

\subsection{Details on the MC simulation of hard helices}
\noindent The position of a helix is defined by the coordinates of its center of mass, while its orientation is specified by 
the unit vectors $\widehat{\mathbf{u}}$, parallel to its axis, and $\widehat{\mathbf{w}}$, parallel to its two-fold symmetry axis (see Figure \ref{fig:helix}).
 The combination of the two unit vectors provides the equivalent information of the three Euler angles $\omega=(\theta,\phi,\psi)$.

Unlike spherical particles, uniaxial objects, such as spherocylinders, require rotational moves of their $C_\infty$ axis ($\widehat{\mathbf{u}}$) 
in addition to the translation of their center of mass. Biaxial particles require an additional rotation of the $\widehat{\mathbf{w}}$  transversal axis. 
There are several ways of accomplishing  this task. One possibility,
due to Barker and Watts \cite{AllenTildesley}, consists on selecting with equal probability one of the three axes of the computational box,
and perform a random rotation of $\widehat{\mathbf{u}}$ around that axis, supplemented by an additional random rotation of $\widehat{\mathbf{w}}$ around $\widehat{\mathbf{u}}$ in the case of biaxial particles. 
A convenient (and equivalent) alternative 
hinges on the use of quaternions, thus avoiding a repeated use of trigonometric equations that are typically time consuming. 
A quaternion can be defined as the unit vector in four dimensional space  $Q \equiv (q_0, q_1, q_2, q_3)$ with $q_0^2+q_1^2+q_2^2+q_3^2 = 1$. Quaternion offers an efficient way to generate uniform random vectors  
on the four dimensional unit sphere \cite{Vesely82}. The one-to-one correspondence between the Euler angle and the quaternion
representation is given by
\begin{align*}
q_0 =& \cos\left(\frac{\theta}{2}\right) \cos\left(\frac{\phi+\psi}{2}\right) \\
q_1 =& \sin\left(\frac{\theta}{2}\right) \cos\left(\frac{\phi-\psi}{2}\right) \\
q_2 =& \sin\left(\frac{\theta}{2}\right) \sin\left(\frac{\phi-\psi}{2}\right)  \\
q_3 =& \cos\left(\frac{\theta}{2}\right) \sin\left(\frac{\phi+\psi}{2}\right)  \\
\end{align*}
and the rotation matrix defining the rotation is given by
\[ \left( \begin{array}{ccc}
q_0^2+q_1^2-q_2^2-q_3^2 & 2(q_1q_2-q_0q_3) & 2(q_1q_3+q_0q_2) \\
2(q_1q_2+q_0q_3) & q_0^2-q_1^2+q_2^2-q_3^2 & 2(q_2q_3-q_0q_1) \\
2(q_1q_3-q_0q_2) & 2(q_2q_3+q_0q_1) &  q_0^2-q_1^2-q_2^2+q_3^2\end{array} \right)\]

Finally, in a NPT calculation, a volume move is required, where one or more lenghts of the box are randomly changed so volume is increased or decreased. This is a global move, involving the whole system, 
and hence should be performed with a relative frequency of $1/N$ with respect to the local translational and rotational moves of a single helix.
On average, we then attempted $N/2$  translations and $N/2$ rotations with equal probability, and one volume move within each MC step. Typical equilibration times were of the order of $10^9$ MC steps, 
followed by production runs
for collecting statistics of the order of $3 \times 10^9$ steps.

Simulations with hard particles consume more computational time compared those with soft particles. This is because of the overlap check for each pair of particles. In our model systems (see Figure \ref{fig:helix}), the distance between two particles cannot be less than the diameter $D$ of a bead. The overlap check in such  non-convex particles needs some tricky algorithms to reduce the computer time. 
In our calculations we first inserted a helix into a suitable spherocylinder, for which a fast test for overlapping is available,
and then tested for possible overlaps at the level of single spheres forming the helix, only in the event of overlapping of the embedding spherocylinders. In this way, the computational time to simulate a fluid of 
hard helices formed by $12$ partially fused spheres and having a fixed contour length of $L/D=12$, is roughly a factor $8$ larger compared with
the corresponding fluid of spherocylinders. 

Most of our simulations were carried out with periodic boundary conditions (PBC)
on a floppy (i.e. shape adaptable) triclinic box, but 
no significant differences were found when using a cuboidal box.
Such conditions are fully compatible with the existence of helical order with small periodicity, comparable with the particle length. 
However they do not  allow in general the development of a cholesteric organization because systems of prohibitively large sizes would be required 
when periodicity is 
much longer than the scale of particle chirality.  
For this reason using PBC in our simulations we could find nematic phases, 
with uniform rather than twisted $\widehat{\mathbf{n}}$ director (i.e. $\cal{P} \rightarrow \infty$). 

In our simulations we used a number of 
particles varying from $900$ to $2000$, always finding rather stable results. 
The value $900$ was then used for most of the simulations.

\section{Onsager (density functional) theory}
\label{sec:onsager}
\noindent This section describes the Onsager theory for 
the isotropic-to-nematic phase transition in a system of hard, slender, rod-like particles \cite{Onsager}.
Beside serving as a basis for all those specific calculations described in the following sections, 
this theory is of general importance for the
whole (soft) condensed matter science.
Onsager formulated it in the forties \cite{Onsager}.  
That is,  in the same period when Kirkwood remarkably conjectured that 
a hard-sphere system, despite the absence of any attractive
interactions, will ultimately exhibit, on increasing density,
a disorder-order transition between a fluid phase and a crystal phase \cite{kirkwood}.
Since no other interactions are operative but hard steep repulsions,
this \emph{ordering} phase transition is purely driven by the sole \emph{entropy}.
That is, at a sufficiently high density, the crystal is thermodynamically stabler
than the fluid because its entropy is larger.
Such a statement was rather counterintuitive and not easily accepted at that time.
Yet, the Kirkwood conjecture was later confirmed in 
the earliest applications of the numerical simulation techniques \cite{Wood,Alder}.
The Onsager theory proceeded, in essence, along the same lines yet, 
exploiting the peculiarity of a system of hard, long and thin, rods, 
it proved  analytically, rather than numerically, 
the entropic origin of the disorder-order transition between 
an isotropic liquid phase and a nematic liquid crystal phase.  
Thanks to these basic works, the concept of \emph{ordering entropy}
progressively gained acceptance and is nowadays well-established.

\begin{figure}[h!]
\centering
\includegraphics[scale=0.5]{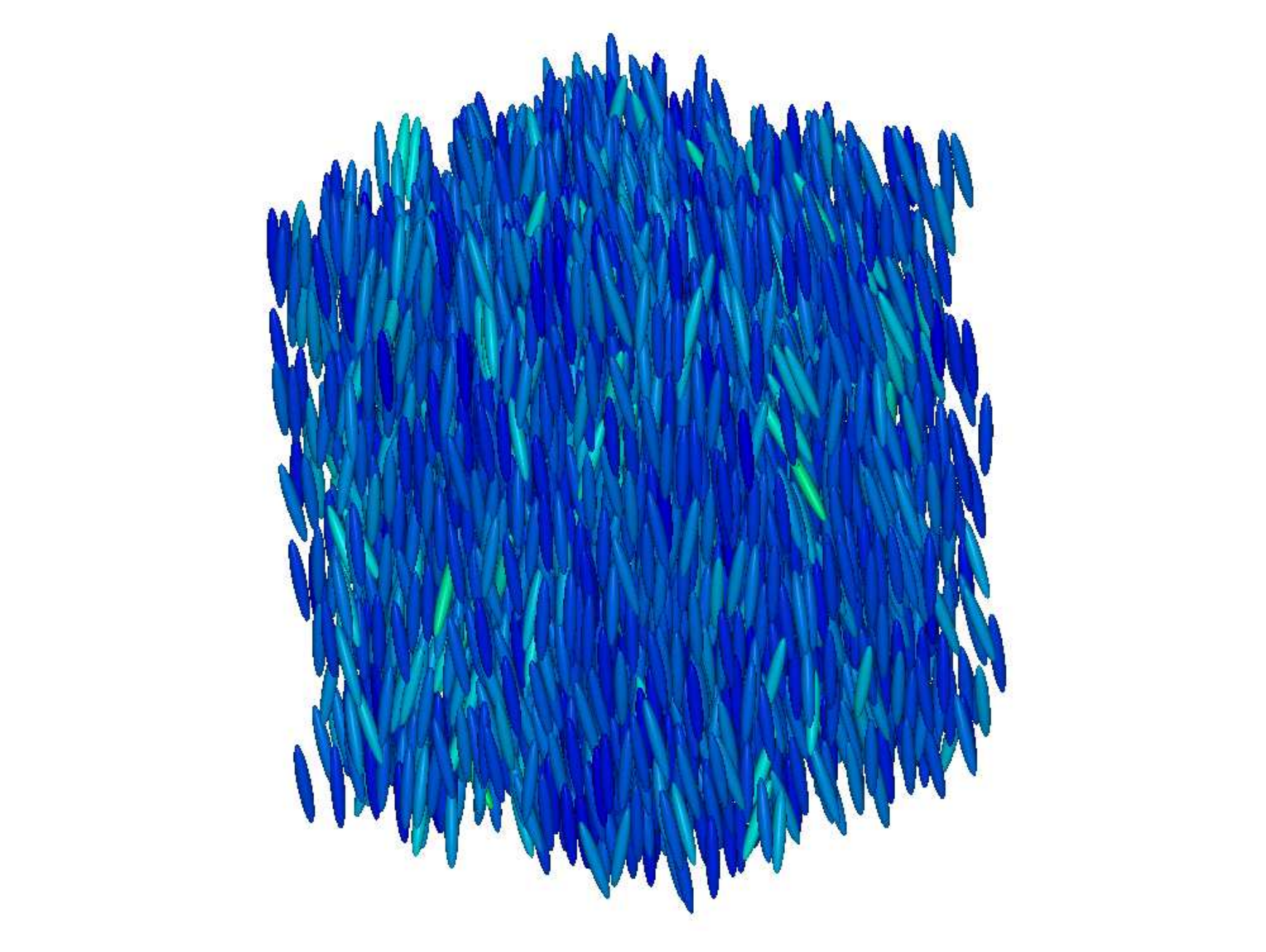}
\caption{Image of a system of hard prolate ellipsoids in the nematic liquid crystal phase.
The image was created with the program QMGA\cite{qmga}.}
\label{fig1}
\end{figure}

Let us suppose to have a system of hard rod-like particles.
For example, the one shown in Figure \ref{fig1},
formed by prolate ellipsoids that 
interact between them only through hard repulsive interactions. 
Looking at this figure, 
the particle centroids appear 
uniformly distributed in space while
the particle main ($C_{\infty}$) axes as roughly aligning along a common direction,
the \emph{director} $\widehat{\mathbf{n}}$: 
this is a snapshot of a system in a nematic liquid crystal phase.  

Hard-particle systems of this sort mimic rather closely 
suspensions of  rod-like colloidal particles 
interacting through short-range repulsive interactions.
In fact, Onsager set out and developed his theory to explain
earlier experimental results on suspensions of this kind,
either of inorganic (mineral) origin,
such as the suspensions made of $\rm{V_2 O_5}$ colloids \cite{zocher} or
organic (biological) origin  
such as those those formed by the tobacco mosaic virus particles \cite{tmv}.
In these experiments, a transition between two phases was observed 
at low concentration of colloids. 
Onsager showed that the transition, involving  
an isotropic liquid phase  and a nematic liquid crystal  phase, 
was due to the highly anisometric shape of the colloidal particles.

One can view a system of $N$ hard, equally sized and rigid, rods, 
uniformly distributed in a container of volume $V$ and free to rotate,
as a (infinite-component) mixture, 
each component being identified by the orientation, $\omega$. 
With the symbol $\omega$ one intends a set of 
variables required to define the orientation of a particle
in the laboratory frame of reference. 
For example, in the case of cylindrically symmetric particles, 
the components of the  unit vector $\widehat{\mathbf u}$ along 
the particle's $C_{\infty}$ axis. 
One can then define a distribution function $f(\omega)$  
such that $N f(\omega)d\omega$ gives 
the number of particles 
whose orientation is comprised within the interval $[\omega, \omega+d\omega]$.
The distribution function $f(\omega)$ is normalised such that $\displaystyle \int d\omega f(\omega) = 1$.

In full analogy with the expression of the entropy of
an ideal multi-component mixture, one can write
the expression of the \emph{ideal} entropy of
the hard-rod system as
\begin{equation}
S_{id}= - N k_{B} \left[\log 
\left ( 
\frac{\rho \Lambda^3_{tr} \Theta_{rot}}{T}
\right ) 
-1 + \int d\omega f(\omega) \log f(\omega)\right],
\label{entro}
\end{equation} 
with $k_B$ the Boltzmann constant, 
$\Lambda_{tr}$ is the thermal wavelength and $\Theta_{rot}$ is the rotational temperature \cite{McQuarrie}. 
The integral in Eq. \eqref{entro} plays the role of a mixing entropy. 

In the aforementioned experiments, 
significant deviations from the ideal behaviour   
were however observed even at low concentration. 
This is due to the hard interactions between the particles, 
whose effect is appreciable even at low concentration 
provided the particles are sufficiently anisometric and
therefore that effect significantly depends on particle orientation.

In a dilute and uniform system of particles interacting with short-range interactions,
the pressure $P$ can be expressed as a \emph{virial} series in
the number density $\rho=N/V$:
\begin{equation}
P= \rho k_B T \left[ 1 + B_2 (T) \rho +B_3 (T) \rho^2 + ... \right], 
\label{virial}
\end{equation}
with $T$ the absolute temperature and the coefficients $B_{\ell}$, $\ell=1,2,...$
termed virial coefficients. 
The first coefficient $B_1$=1 while
the successive coefficients, starting with the second-virial coefficient $B_2$,
depend on particle interactions and $T$.

Originally devised as an empirical way to account for deviations from the ideality,
the virial series was given a statistical-mechanical foundation by
the remarkable work of Mayer and Mayer \cite{mayer}.
They showed that the virial coefficients can be written 
as integrals, of rapidly increasing complexity as the order $\ell$ increases, 
whose integrands depend on the Mayer function $M$. The latter is defined as:
\begin{equation}
\label{eq:mayer}
\displaystyle M\left(\mathbf{R}_i, \omega_i, \mathbf{R}_j, \omega_j \right) = 
\mathsf{e}^{-\frac{u\left(\mathbf{R}_j-\mathbf{R}_i, \omega_i, \omega_j \right)}{k_B T}} - 1,
\end{equation}
with $\mathbf{R}_i$ the vector giving the position of particle $i$'s centroid
in the laboratory frame of reference and 
$u\left(\mathbf{R}_j-\mathbf{R}_i, \omega_i, \omega_j \right)$ the potential energy of interaction between particles $i$ and $j$. 
In particular, the expression for the leading second-virial coefficient is
\begin{equation}
B_2(\omega_i,\omega_j|T) = - \frac{1}{2} \int d\omega_i f(\omega_i) \int d\omega_j  f(\omega_j) \int d\mathbf{R} M\left(\mathbf{R},\omega_i,\omega_j\right).
\end{equation} 
Observe that for hard interactions, the Mayer function takes on two values only:
-1 if particles overlap and 0 if they do not. 
The dependence on temperature drops off:
a hard-particle system is athermal.
The second-virial coefficient can thus be re-written as:
\begin{equation}
B_2(\omega_i,\omega_j) = \frac{1}{2} \int d\omega_i f(\omega_i) \int d\omega_j  f(\omega_j) 
v(\omega_i, \omega_j),
\end{equation}
with $v(\omega_i,\omega_j)$ the \emph{volume} that particle $i$ \emph{excludes} to particle $j$.
The expressions for the next virial coefficients are increasingly more complicated 
to such an extent that 
the computational burden to evaluate them rapidly becomes prohibitively heavy
even for hard spheres, the simplest hard-particle model.
Thus, exact summation of a virial series is in general precluded and 
one has to resort to approximations 
whose reliability should always be tested. 
The system of hard slender rods is however an exception in this respect.

In fact, Onsager made the crucial observation that, 
as long as the particles are needle-like,
the terms of order $\ell$ higher than two can be neglected in Eq. \eqref{virial}.
In this case, the system's \emph{excess}  free energy $\mathsf{F}^{ex}$,
obtained from the non-ideal part of equation of state by 
standard thermodynamic integration, results:
\begin{equation}
\mathsf{F}^{ex}= {\rm{N}} k_B T \frac{\rho}{2} \int d\omega  f(\omega) \int d\omega^{'} 
 f(\omega^{'}) v(\omega, \omega^{'}).
\label{freenergy}
\end{equation}  

Summing up Eqs. \eqref{entro} and \eqref{freenergy} provides the system's total free energy $\mathsf{F}$:
\begin{equation}
\frac{\mathsf{F}}{{\rm N} k_B T} = \log
\left ( 
\frac{\rho \Lambda^3_{tr} \Theta_{rot}}{T}
\right ) 
 - 1 + \int d\omega f(\omega) \log f(\omega) +
\frac{\rho}{2} \int d\omega f(\omega) \int  d\omega^{'} 
 f(\omega^{'}) v(\omega, \omega^{'}).
\label{freetotal}
\end{equation}

Observe that the free energy is a \emph{functional} of the  
distribution function $f(\omega)$: the Onsager theory is a precursor
of modern density functional theories.
The thermodynamic stable state of the system at a given  $\rho$ corresponds
to that specific distribution function $f(\omega)$ that minimises the free energy.
If $f(\omega)$ is constant, that is
the particle orientations are uniformly distributed, 
the orientational entropy term is maximised.
Conversely, the density-dependent 
excluded-volume term is minimal when particles align.
It is from the competition between these two terms that
the isotropic-to-nematic phase transition arises 
at sufficiently high density.

The rigorous minimisation of the free energy functional under the constraint of
distribution function normalisation leads
to the following integral equation:
\begin{equation}
\log K f(\omega) = - \rho \int d\omega^{'} f(\omega^{'}) v(\omega, \omega^{'}), 
\label{integeq}
\end{equation}
with $K$ a constant whose value is determined by ensuring the normalisation condition.
Its analytic solution is not available.
Onsager by-passed this difficulty by resorting to an approximate trial function $f(\omega)$
whose form can be regulated by the value of a single parameter $\alpha$ and minimised
directly the free energy (Eq. \eqref{freetotal}) with respect to $\alpha$. 
This procedure can be extended to functional forms containing an arbitrarily large number of variational parameters. A common choice is that of using a truncated expansion of $f(\omega)$ on a set of orthonormal basis functions. 
The expansion coefficient are the orientational order parameters.
The integral Eq. \eqref{integeq} can nonetheless be solved  
by an iterative algorithm that, starting from a guessed orientational distribution function, ends whenever 
self-consistency is satisfied within a given tolerance.

One can thus see that, beside the isotropic-liquid  solution of Eq. \eqref{integeq}, corresponding to a constant $f(\omega)$ and always present for any value of $\rho$,
an additional anisotropic-liquid solution emerges 
provided the value of $\rho$ is sufficiently high \cite{kayser,herzfeld}. 
Inserting the isotropic and anisotropic solutions into the expression for the pressure,
the equation of state for the isotropic phase and nematic phase
can respectively be obtained. 
From standard thermodynamics, the chemical potential expressions
for the two branches can also be obtained.
Equating pressure and chemical potential, the value of $\rho$ at which
the isotropic phase and nematic phase are in equilibrium can be thus determined.
The Onsager theory predicts the isotropic-to-nematic phase transition to be
first-order in agreement with experimental evidence.
Provided the experimental rod-like colloids have 
a sufficiently large aspect ratio,
like the tobacco mosaic virus particles,
the agreement is essentially quantitative \cite{fraden}.
Definite confirmation of the validity of 
the Onsager theory came from numerical experiments on model hard-particle systems
such as that shown in Figure \ref{fig1}.

Deviations from Onsager predictions increase with decreasing aspect ratio, 
because of the progressively higher contribution of virial coefficients beyond second order \cite{Frenkel}.
In earlier calculations approximate values of
$B_3$ were obtained \cite{Straley73,Mulder}, and recently 
the  eighth virial level has been reached \cite{Masters}.  
Surprisingly good agreement between theory and simulations (in systems
of hard rod-like particle) could be obtained  
using the Parsons-Lee expression \cite{ParsonsLee}, 
where many-body effects are introduced in an approximate way, 
by multiplying the Carnahan-Starling  form of the excess free energy of hard spheres  \cite{Carnahan} by the second virial coefficient specific of the particles under investigation \cite{allen}. 
So, the excess free energy reads:
\begin{equation}
\mathsf{F}^{ex}= {\rm{N}} k_B T G(\eta) \frac{\rho}{2} \int d\omega f(\omega) \int  d\omega^{'} 
 f(\omega^{'}) v(\omega, \omega^{'})
\label{eq:freenergyPL}
\end{equation}  
where $G(\eta)=(4 -3\eta)/4(1-\eta)^2$.  Here $\eta=\rho v_{0}$ is the volume fraction, with $v_0$ being the volume of a particle.
In ref. \cite{Frezza13} we used the Parsons-Lee approximation to investigate the isotropic-nematic transition of hard helical particles and compare with the results of MC simulations (see Figure \ref{fig:P2_result}). 

The Onsager theory forms the basis of all successive developments in the field.
The case of hard helices is no exception in this respect. 
\section{Onsager-like theory for the cholesteric and screw-nematic phases}
\label{sec:DFThelical}
\noindent  
We have considered two kinds of helically modulated nematic phases: the cholesteric and the screw-nematic.
In the former  the nematic director $\widehat{\mathbf n}$, 
i.e. the average alignment axis of the helical axes  
($\widehat{\mathbf u}$), spirals around a perpendicular axis ($Y$ axis of the laboratory frame in Figure \ref{fig:helix}). 
In the screw-nematic (N$_s^{\ast}$) phase it is the secondary polar director $\widehat{\mathbf c}$, 
i.e. the average alignment axis  of the   $\widehat{\mathbf w}$ axes of helices that spirals around  
$\widehat{\mathbf n}$ ($Z$ axis of the laboratory frame in Figure \ref{fig:helix}).
For such phases the Onsager theory outlined in the previous section has to be generalized to include explicit dependence 
upon positional variables. In both cases  a single positional variable is needed, 
i.e. the position along the helical axis, because there is translational homogeneity perpendicularly to such an axis.
In the case of a helically modulated phase with the helical axis along the direction $h$ and pitch $\cal P$  
we can introduce the orientational distribution function at the $h$ position, $f(\omega | h)$, 
which has  periodicity $\cal P$ too,  and is normalized such that $\int d\omega f(\omega| h) = 1$, 
irrespective of $h$. 
By retaining only the second and third terms  in the virial expansion,
the excess Helmholtz free energy of such a system is given by:
\begin{eqnarray}
\frac{\mathsf{F}^{ex}}{V k_B T} & =& 
\frac{\rho^2}{2 {\cal P}}  \int_0^{\cal P}   d h \int d\omega f(\omega | h) \int d h^{\prime} \int d\omega^{\prime} 
  f(\omega^\prime | h^\prime)) \nonumber \\
&& \biggl[ a_{excl}( h, \omega, h^{\prime}, \omega^{\prime}) + 
\frac{\rho}{3} \int d h^{\prime \prime} \int d \omega^{\prime \prime} 
f(\omega^{\prime \prime} | h^{\prime \prime}), \omega^{\prime \prime})  
a_3(h, \omega, h^{\prime}, \omega^{\prime}, h^{\prime \prime}, \omega^{\prime \prime}) \biggr] ,
\label{eq:uno}
\end{eqnarray}
where the functions $a_{excl}( h, \omega, h^{\prime}, \omega^{\prime})$ and
$ a_3(h, \omega, h^{\prime}, \omega^{\prime}, h^{\prime \prime}, \omega^{\prime \prime}) $ 
have been introduced.
The first is given by:
\begin{equation}
\label{eq:2}
a_{excl}( h, \omega, h^{\prime}, \omega^{\prime})=
-\int d{\bf R}_\perp^{\prime} M(0,h,0,\omega,{\bf R}_\perp^{\prime},h^{\prime},\omega^{\prime})
\end{equation}
where $M$ is the Mayer function, Eq. \eqref{eq:mayer} and ${\bf R}_\perp$ is the vector position in the plane perpendicular to $h$.
$a_{excl}$  is interpreted as the area of the surface obtained by cutting 
with a plane perpendicular to the helical axis at position $h^{\prime}$
the volume excluded to a particle with orientation $\omega^{\prime}$ by
a particle at position $h$ with orientation $\omega$ \cite{Cinacchi04}. 
The second function in Eq. \eqref{eq:uno} is given by: 
\begin{eqnarray}
  \label{eq:new}
&&a_3( h, \omega, h^{\prime}, \omega^{\prime}, h^{\prime \prime}, \omega^{\prime \prime})=
-\int  d{\bf R}_\perp^{\prime}  \int d{\bf R}_\perp^{\prime \prime} 
M(0,h,\omega,{\bf R}_\perp^{\prime},h^{\prime},\omega^{\prime}) \nonumber \\
&& M(0,h,\omega,{\bf R}_\perp^{\prime \prime},h^{\prime \prime},\omega^{\prime \prime})
M({\bf R}_\perp^{\prime},h^{\prime},\omega^{\prime},{\bf R}_\perp^{\prime \prime},h^{\prime \prime},\omega^{\prime \prime}).
\nonumber \\
\end{eqnarray}
Since  $a_{excl}$ and $a_3$  actually depend on the differences 
$\zeta^{\prime}=h^{\prime}-h$ and $\zeta^{\prime \prime} = h^{\prime \prime} - h$,
Equation(\ref{eq:new}) can be rewritten as:
\begin{eqnarray}
\frac{\mathsf{F}^{ex}}{k_B T V}  &=& 
\frac{\rho^2}{2}  \int  d\omega   f(\omega| h=0) 
\int d \zeta^{\prime} \int d\omega^{\prime}   f(\omega^\prime |\zeta ^\prime)
\biggl[a_{excl}( \omega, \zeta^{\prime}, \omega^{\prime}) \nonumber \\
&&+ \frac{\rho}{3} \int d \zeta^{\prime \prime} \int d \omega^{\prime \prime} 
 f(\omega^{\prime \prime} | \zeta^{\prime \prime}) 
a_3(\zeta^{\prime},\omega,\omega^{\prime}, \zeta^{\prime \prime}, \omega^{\prime \prime}) 
\biggr]. 
\label{neat}
\end{eqnarray}
If the Parsons-Lee approximation is used, the following expression is obtained
\begin{equation}
\frac{\mathsf{F}^{ex}}{k_B T V}  = 
\frac{\rho^2}{2}  G(\eta)  \int  d\omega   f(\omega| h=0) 
\int d \zeta^{\prime} \int d\omega^{\prime}   f(\omega^\prime |\zeta ^\prime) a_{excl}( \omega, \zeta^{\prime}, \omega^{\prime}) 
\label{freenergyPL2}
\end{equation}  
For a uniform nematic phase $f(\omega|h=0)=f(\omega)$, 
and if the expansion is truncated at the second virial term, 
Eq. \eqref{eq:freenergyPL} is recovered. 
The equations above are equally valid for the cholesteric and screw-like nematic phase.
In the former case $h=Y$ and ${\cal P}$ is the cholesteric pitch;
in the latter case $h=Z$ and ${\cal P}$ coincides with $p$ the pitch of a single helix  

The free energy of the cholesteric and the screw-nematic phase 
bears a dependence upon the helical pitch $\cal P$, introduced by the orientational distribution function $f(\omega|\zeta)$. 
Thus, determination of the equilibrium state at a given density requires minimization of the free energy with respect to this parameter, in addition to the other quantities that define the orientational distribution function.
The numerical handling of the free energy of helically modulated nematic phases is significantly more demanding than for the uniform case, not only because of the positional dependence which explicitly appears in Eqs. \eqref{eq:2}, but also because of the inherent biaxiality of such modulated phases. Similar approach has been recently implemented for the twist-bend nematic phase \cite{GF15}. 
When dealing with the cholesteric phase,  some simplifying assumptions can be used, taking advantage of the long scale of the twist deformation, compared to scale of inter-particle interactions.
A first consequence of the long cholesteric pitch is that the biaxiality is very small, so as a first approximation it can be neglected. 
This was done in Refs.  \cite{Belli14,DussiJCP}, where a Parson-Lee treatment was proposed, based on expressions of the free energy in terms of uniaxial order parameters and  the wavenumber $q$.
The Straley approach can also be adopted  \cite{Straley}, where  further assumptions are made, i.e. 
that  $f(\omega)$ in the N$^\ast$ phase is identical to that in the undeformed N phase at the same density, but with respect to a director that rotates in helical way, and that the $q$ dependence  of the excess free energy can be accounted for by the following truncated Taylor expansion:
\begin{equation}
\mathsf{F}^\text{ex}=\mathsf{F}^\text{ex}|_{q=0}+\left . \frac{d \mathsf{F}^\text{ex}}{d q}\right |_{q=0} \, q 
+\frac{1}{2}\left . \frac{d^2 \mathsf{F}^\text{ex}}{d q^2} \right |_{q=0}  \, q^2. 
\label{eq:aexp}
\end{equation}
Here the first term on the rhs is the free energy density of the undeformed nematic phase. 
So, the difference between the free energy of the twisted and the uniform nematic phase, $\mathsf{F}^\text{ex}-\mathsf{F}^\text{ex}|_{q=0}$, 
can be compared with the elastic energy of the cholesteric phase, which according to the continuum elastic theory reads \cite{deGennesbook}:
\begin{equation}
\frac{\mathsf{F}_{el}}{V} = k_2 \,q + \frac{1}{2} K_{22} \, q^2 
\label{Ieq:eq1}
\end{equation}
where $K_{22}$ is the \emph{twist elastic constant} and $k_2$ is the \emph{chiral strength}. The former is a positive quantity accounting for the energetic cost of twist deformations, and has  typical values  of the order of  piconewtons. 
The chiral strength $k_2$ is a pseudoscalar, which has opposite values for systems which are the mirror image of each other and vanishes in the achiral nematic phase.
The equilibrium pitch of a cholesteric phase is obtained by minimization of the elastic energy and is given by: 
\begin{equation}
{\cal P}=-2 \pi K_{22}/k_2;
\label{Ieq:eq2}
\end{equation}
it becomes infinitely long in achiral nematic liquid crystals. 

Comparing Eqs. \eqref{eq:aexp} and \eqref{Ieq:eq1} we can write the microscopic expression for the chiral strength and the twist elastic constant:
\begin{eqnarray}
k_2 =& \displaystyle \left . \frac{d \mathsf{F}^\text{ex}}{d q}\right |_{q=0} \label{Ieq:eq3} \\
K_{22} =& \displaystyle \left . \frac{d^2 \mathsf{F}^\text{ex}}{d q^2}\right |_{q=0}. 
\label{Ieq:eq4}
\end{eqnarray} 
These two quantities behave differently under inversion: the former is invariant (scalar), whereas the latter changes its sign (pseudoscalar). Thus,  for chiral particles that are the mirror image of each other (enantiomers), the free energy of the undeformed nematic phase and the twist elastic constant are identical, whereas $k_2$ takes opposite values.

\begin{figure}[h!]
\centering
\includegraphics[
width=8cm]{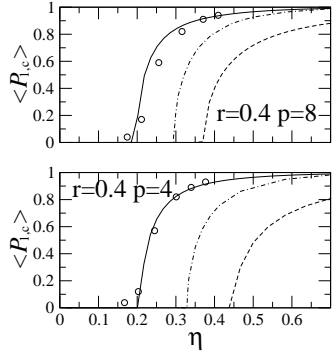}
\caption{Comparison between numerical simulations (circles) and Onsager-like theory for perfectly parallel helices, in the plane
screw-nematic order parameter-volume fraction.
Lines are theory results using the II virial approximation (dashed),
the II virial  approximation with Parson-Lee correction (dash-dotted), and the III virial approximation (solid).}
\label{fig_giorgio}
\end{figure}

In our calculations for the cholesteric phase of hard helices  \cite{Frezza14} we followed Straley approach, along the lines described in ref. \cite{Tombolato05}. In the case of the screw-nematic phase, we adopted the simplifying assumption of perfectly parallel helical axes; thus helices had only translational and azimuthal freedom \cite{Kolli14b}. This could be justified by the high degree of orientational order that characterizes the screw-nematic phase. However, the high value of density at which the N-N$^\ast_S$ transition occurs 
called for considering also the explicit contribution of the third-virial terms.
Indeed, it was seen that their inclusion is needed to achieve
a nearly quantitative agreement with corresponding
numerical simulation data, as illustrated in Figure \ref{fig_giorgio}.
The generalisation of these results to systems of freely rotating
hard helices is planned for the near future. 
While conceptually clear (Eqs. 6.1-6.4), 
it is expected to require a considerable numerical
effort not only because of 
the complexity of a helically modulated phase but also 
because of the incorporation of the third-virial term contribution. 
This task will none the less be rewarding in that 
it will allow to treat both cholesteric and screw-like nematic
phases at the same level of theory and study
accurately their mutual phase transition and the transition between these and the isotropic phase.
\section{Order parameters and correlation functions}
\label{sec:OP}
\noindent Different phases are characterized by appropriate order parameters capturing
their distinctive symmetry and the corresponding degree of ordering, and
by suitable correlation functions. In the following we will present the order parameters and correlation functions that have been used to characterize the different phases formed by helical particles.
Some of the latter are conveniently resolved along the $\widehat{\mathbf{n}}$ director ($\parallel$) and perpendicular to it ($\perp$), according to the following decoupling of the inter-particle vector 
$\mathbf{R}_{ij} = \mathbf{R}_{i}-\mathbf{R}_{j}$: 
\begin{eqnarray}
\mathbf{R}_{ij}= \mathbf{R}_{ij}^{\parallel}+  \mathbf{R}_{ij}^{\perp}=\left(\mathbf{R}_{ij} \cdot \widehat{\mathbf{n}} \right) \widehat{\mathbf{n}}+
\left \vert \mathbf{R}_{ij} \times \widehat{\mathbf{n}} \right \vert \widehat{\mathbf{R}}_{ij}^{\perp}
\label{eq5b}
\end{eqnarray}
 
\subsection{Nematic order parameter $\langle P_2 \rangle$}
\noindent Nematic ordering, i.e. the degree of (non-polar) alignment of the  $\widehat{\mathbf{u}}$ axes of helices with respect to the $\widehat{\mathbf{n}}$ director is quantified by the order parameter:
\begin{equation}
\langle P_2\rangle = \left \langle \frac{3}{2} \cos^2\theta - \frac{1}{2} \right \rangle 
\label{eq:nem_op}
\end{equation}
where  $P_2$ is the second Legendre polynomial and $\cos \theta= \widehat{\mathbf{u}} \cdot \widehat{\mathbf{n}}$. Here and henceforth angular brackets denote ensemble averages. 
$\langle P_2\rangle$ is zero in the isotropic phase and nonzero in the nematic phase. It can be nonzero also in other more ordered phases, in which case there are other additional order parameters.
In a simulation, average values are obtained as mean values over the trajectory; so, $\langle P_2\rangle$  would be obtained as the mean of the following quantities, calculated for each trajectory frame: 
\begin{equation}
 (P_2) _{\rm{frame}} 
                               = \frac{1}{N} \sum_{i=1}^{N} { \frac{3}{2}(\widehat{\mathbf{n}} \cdot \widehat{\mathbf{u}}_{i})^2 - 1} 
\label{eq:nem_op2}
\end{equation}
where the sum is over all particles.
However one does not know at the outset the orientation of the director $\hat{\mathbf{n}}$, as particles can align in principle along any direction in space.
A convenient approach is to use the second rank Veilliard-Baron tensor \cite{Veilliard} 
\begin{equation}
Q_{\alpha \beta} =\frac{1}{N} \sum_{i=1}^N  \left(\frac{3}{2} u_i^\alpha u_i^\beta - \frac{\delta_{\alpha \beta}}{2}\right)
\end{equation}
where  
$\alpha, \beta = x, y, z$ 
are indices referring to the laboratory frame, and $u_i^\alpha$ is the component of the unit vector $\widehat{\mathbf{u}}$ of particle $i$ along $\alpha$. The $\mathbf{Q}$- matrix represents a symmetric, 
traceless second rank tensor. The three eigenvalues $\lambda_1, \lambda_2, \lambda_3$ of Q-matrix are a measure of nematic order in the three orthogonal directions defined by the corresponding eigenvectors. 
In case of uniaxial nematics, $-2\lambda_1 = -2\lambda_2 = \lambda_3$. The largest (in absolute value) eigenvalue, $\lambda_3$, is the nematic order parameter $\langle P_2 \rangle$ and the corresponding 
eigenvector is the nematic director $\widehat{\mathbf{n}}$. 

\begin{figure}
\centering
\includegraphics[width=10cm]{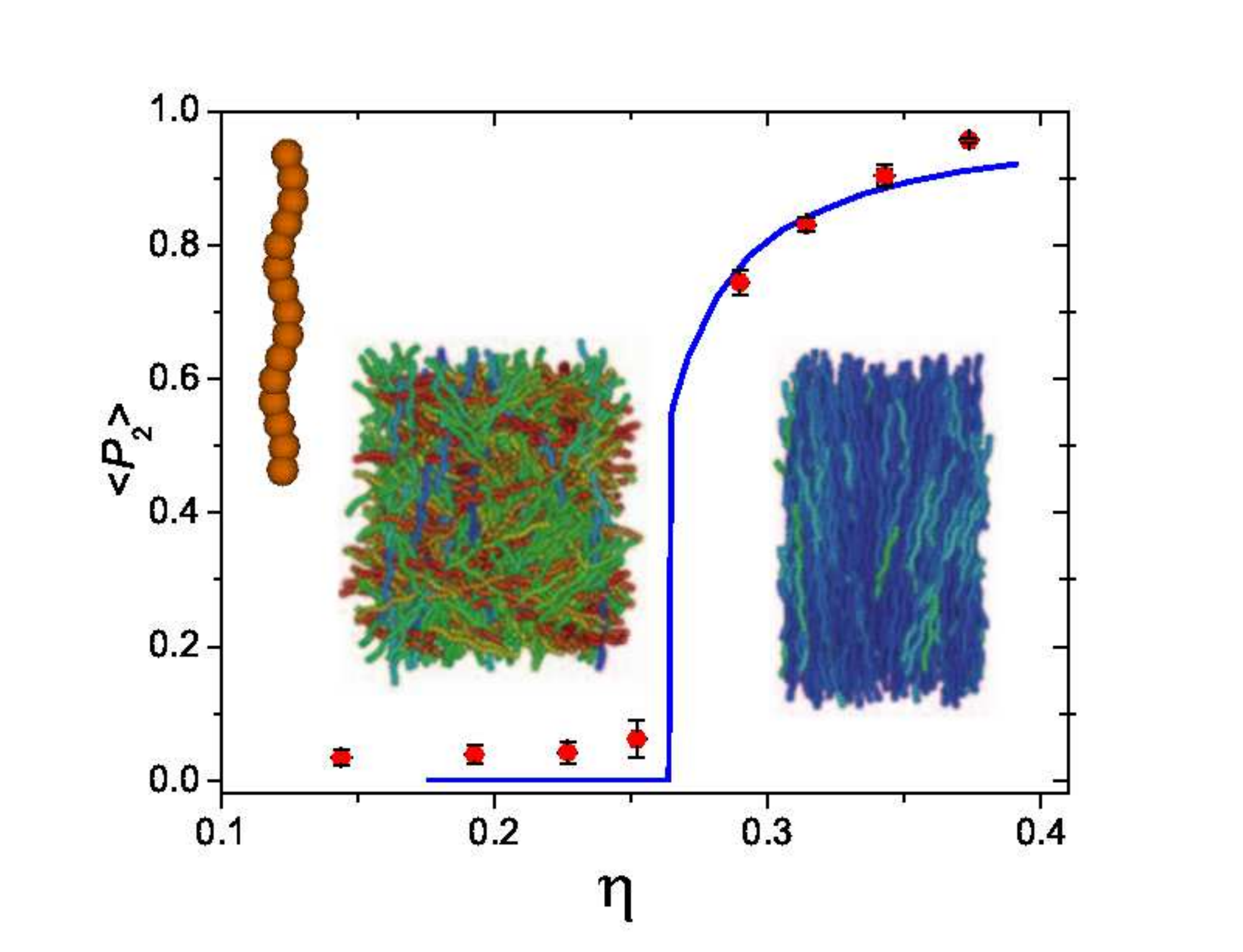}
\caption{The nematic order parameter, $\langle P_2 \rangle$, as a function of the volume fraction $\eta$ for helices with radius $r=0.2$ and pitch $p=4$, whose morphology is displayed in the inset. 
The plot shows the results of NPT-MC simulations (filled circles) and those  of Onsager theory 
with Parsons-Lee correction, see Eq. \eqref{eq:freenergyPL}
(line). Representative snapshots from simulations in the isotropic (left) and in the nematic phase (right) are shown. }
\label{fig:P2_result}
\end{figure}
Figure \ref{fig:P2_result} shows the order parameter  $\langle P_2\rangle$ as a function of the volume fraction $\eta$, for helices with radius $r=0.2$ and pitch $p=4$. We can see the typical behavior, with $\langle P_2 \rangle$ close to zero  at low density (isotropic phase) and  then jumping to a high value beyond a given density. This jump is a signature of the first-order isotropic-to-nematic transition. The fact that $\langle P_2 \rangle$ does not exactly vanish in the isotropic phase is a finite-size effect. Likewise, we find $\lambda_1 \approx \lambda_2$ within a $5\%$ discrepancy, which however can be taken as compatible with the uniaxial character of the conventional N phase. 

\subsection{Screw-like nematic order parameter}
\noindent The phase diagram of helices presents phases that, unlike the conventional nematic, are characterized by screw-like order. This is quantified by the order parameter:
\begin{equation}
\left \langle P_{1,c} \right \rangle = \left \langle \widehat{\mathbf{w}} \cdot \widehat{\mathbf{c}} \right \rangle
\label{sec6:eq5}
\end{equation}
which measures the average alignment along a common direction ($\widehat{\mathbf{c}}$) of the 
two-fold symmetry axes ($\hat{\mathbf{w}}$) of helices.
In phases with screw-like ordering the $\widehat{\mathbf{c}}$ director rotates in helical way around the $\widehat{\mathbf{n}}$ director, with periodicity $\cal P$ equal to the  pitch $p$ 
of the helical particles. 
To calculate $\left \langle P_{1,c} \right \rangle$ from simulations
it is expedient to perform a preliminary
rotation of $-2\pi (\mathbf{R}_{i}\cdot \widehat{\mathbf{n}})/p$ around $\widehat{\mathbf{n}}$ of the particle coordinates. 
The $\widehat{\mathbf{c}}$ director and the  $\left \langle P_{1,c} \right \rangle$ order parameter are then determined for the untwisted structure, following the same procedure used for the calculation of the nematic order parameter.

In smectic phases there is also the possibility of a in-layer transversal polar order without correlation between layers.
Even in this case an order parameter defined according to  Eq. \eqref{sec6:eq5} can be used, 
but the average  has to be meant 
over all particles belonging to individual layers.
To stress this point and distinguish this case from screw-like order, we will use for it the symbol $\langle P_{1,c}^{\rm{layer}} \rangle$.

\begin{figure}
\centering
\includegraphics[width=6cm]{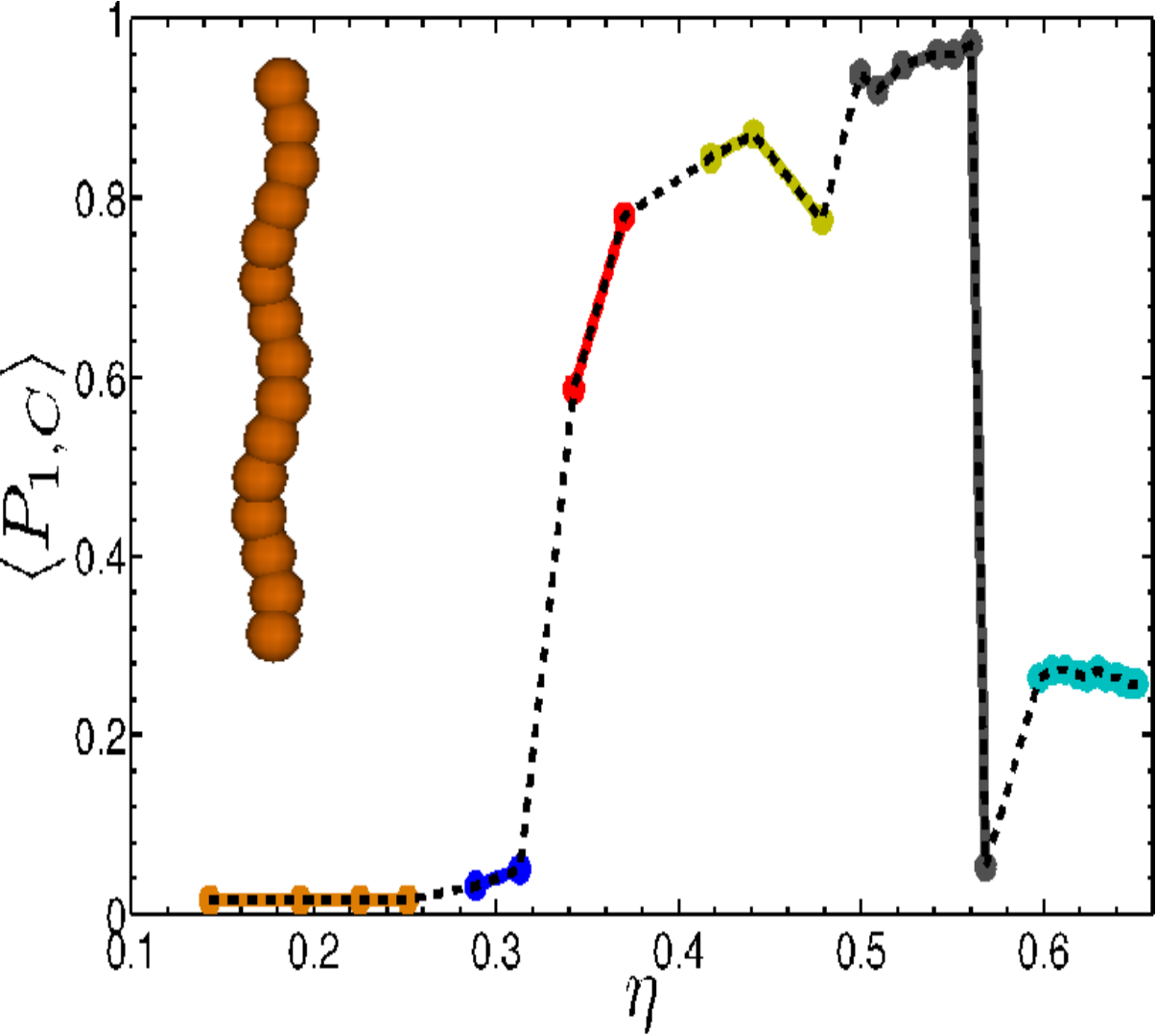}
\caption{Polar order parameter, $\langle P_{1,c} \rangle$, as a function of the volume fraction $\eta$ 
 for helices with radius $r=0.2$ and pitch $p=4$  (see inset).
 The symbols show the results of NPT-MC simulations, lines are meant as a guide to eye.
 Colors are used to distinguish different phases: I (orange), N (blue), N$^\ast_S$ (red), Sm$^\ast_{A,S}$ (yellow), Sm$^\ast_{B,S}$ (gray), C (cyan). Phase labeling is reported in Table I; C is a high density compact phase.}
\label{fig:P1c_result}
\end{figure}

Figure \ref{fig:P1c_result} shows the polar order parameter, $\langle P_{1,c} \rangle$
 as a function of the volume fraction, $\eta$, for helices with radius $r=0.2$ and pitch $p=4$. 
Comparing with the plot of the nematic order parameter, $\langle P_{2} \rangle$, 
calculated for the same helices (see Figure \ref{fig:P2_result}), we can see 
that $\langle P_{1,c} \rangle$ 
is negligibly small in both the isotropic and the nematic phase, 
but  at higher density it raises, reaching values close to 1. 
The different colors in Fig.  \ref{fig:P1c_result} refer to different phases
 in which $\langle P_{1,c} \rangle$ is different from zero; to distinguish between 
 them additional order parameters have to be introduced.
 
\subsection{Smectic order parameter}
\noindent Smectic order can be characterized by the translational order parameter
\begin{eqnarray}
\left \langle \tau_1 \right \rangle &=& \left| \left \langle \exp\left(2\pi \textrm{i} \frac{Z}{d}\right) \right \rangle \right|
\label{eq:tau_1}
\end{eqnarray}
where $Z$ is the position along the director $\widehat{\mathbf{n}}$ and $d$ is the layer spacing.
Since such spacing is not known a priori, $\langle \tau_1 \rangle$ is calculated for different values of $d$. 
 The maximum $\langle \tau_1 \rangle$ value is then taken as the smectic order parameter and the corresponding $d$ value as the layer spacing \cite{Memmer02,Cifelli06}. 

\begin{figure}
\centering
\includegraphics[width=9cm]{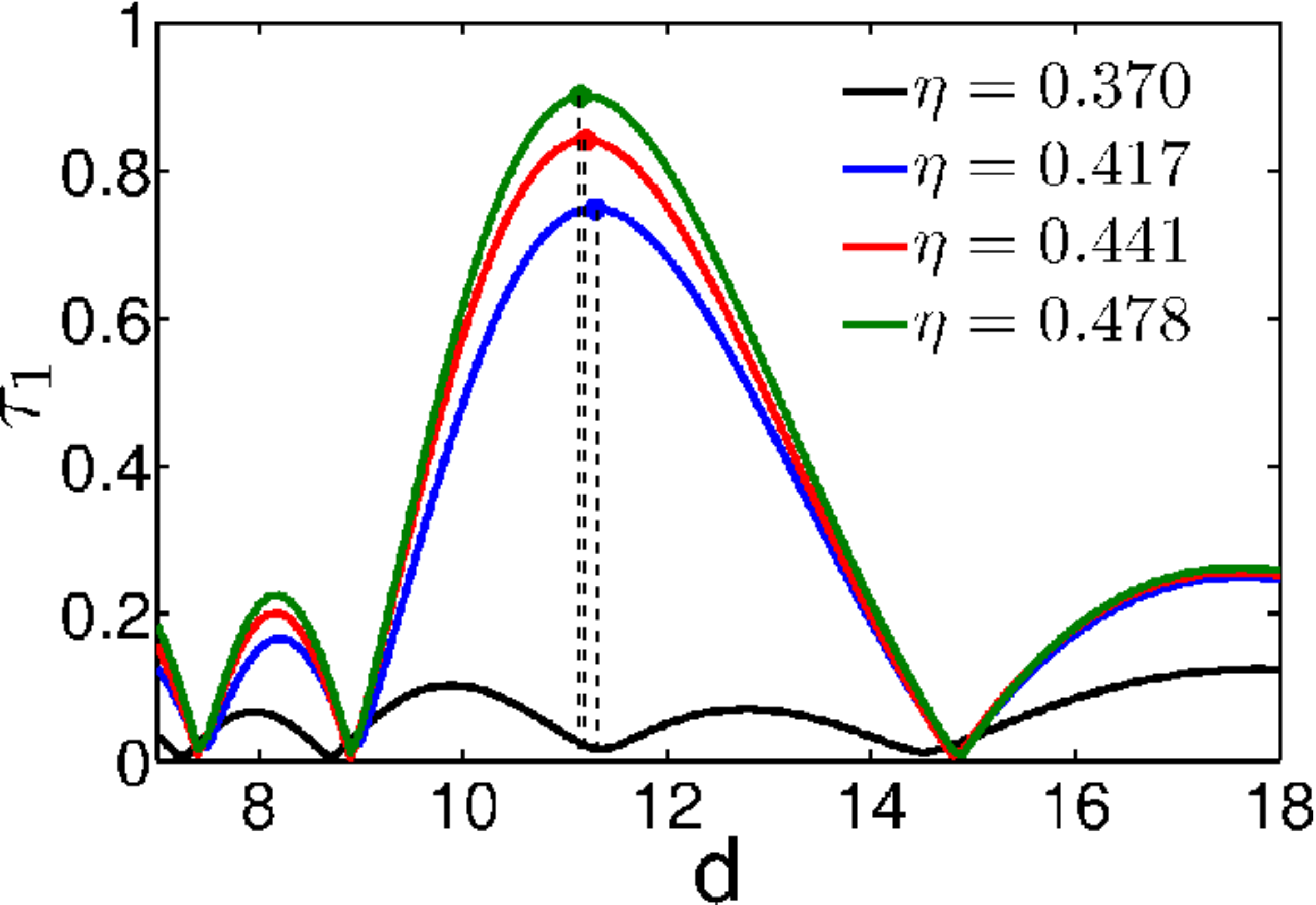}
\caption{The dependence of translational order parameter $\tau_1 \equiv \langle \tau_1 \rangle$ on layer spacing for helices having $r = 0.2$, $p = 4$ on either side of N$_S^*$ - Sm$^{\ast}_{A,S}$ transition. The maximum value of $\tau_1$ is taken as the smectic order parameter.}
\label{fig:smectic_d}
\end{figure}

The order parameter $\langle \tau_1 \rangle$
is close to unity in case of perfect layering and close to zero for non-layered phases, like the I, N and N$_S^*$. 
A representative example is reported in Figure \ref{fig:smectic_d}, where $\langle \tau_1 \rangle$ is  plotted as a function of the layer spacing $d$ calculated at different $\eta$ values for helices 
with $r = 0.2$ and $p = 4$.  The black line in the figure, which is always close to zero, corresponds to the N$_S^*$ phase at $\eta = 0.370$. The blue, red and green lines, showing peaks 
at $d \approx 11.5$, correspond to smectic phases and are in the order of increasing $\eta$. The maximum values of peaks are their corresponding smectic order parameters. With increase in $\eta$, 
the order parameter increases and corresponding layer spacing shows a small tendency to decrease. 
\subsection{Hexatic order parameter}
\noindent The $\langle \tau_1 \rangle$ order parameter captures the onset of smectic ordering but is unable to distinguish between the
smectic B phase, where there is  six-fold bond-orientational (hexatic)  ordering, and other phases, like the smectic A, where there is no in-plane ordering.
This discrimination can however be accomplished by calculating the hexatic order parameter $\langle \psi_6 \rangle$ \cite{Memmer02,Cinacchi08} 
\begin{eqnarray}
\left \langle \psi_6 \right \rangle &=& \left\langle  \frac{1}{N} \sum_{i=1}^{N} \left | \frac{1}{n(i)} \sum_{j=1}^{n(i)} 
\exp{(6i\theta_{ij})}\right |\right\rangle
\label{eq:hexatic}
\end{eqnarray}
Here the inner sum is over all  nearest neighbours of particle $i$ within a single layer, whose number is $n(i)$,
and  $\theta_{ij}$ is the angle between 
the unit vector $\widehat{R}_{ij}^\perp$ defined in Eq. \eqref{eq5b}  and a randomly fixed reference axis in a plane perpendicular to $\widehat{\mathbf{n}}$.  

\begin{figure}
\centering
\includegraphics[width=5cm]{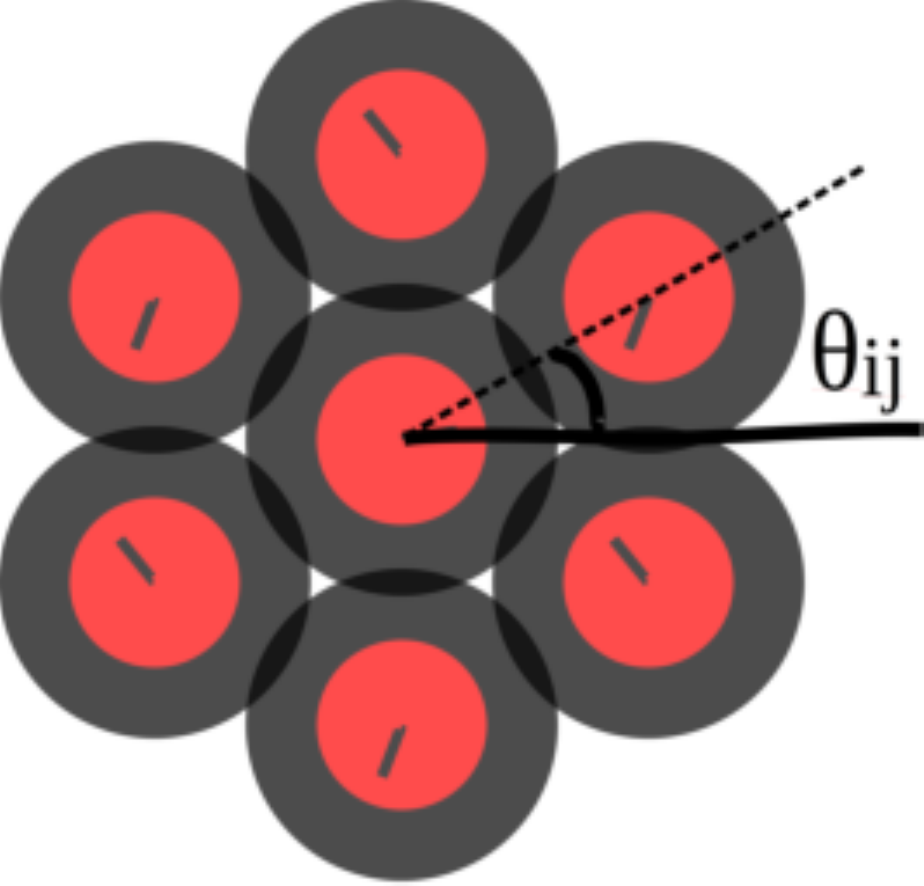}
\caption{Cartoon showing the quantities used to define the hexatic order parameter $\langle \psi_6 \rangle$, Eq. \eqref{eq:hexatic}. The red circles are transversal sections of cylinders enclosing the helices and  the thick short line inside each circle indicates the orientation of the  helix $\widehat{\mathbf{w}}$ vector. The bond angle $\theta_{ij}$ is the angle between a reference axis (thick solid line) in a plane perpendicular to the director $\widehat{\mathbf{n}}$, and the  projection of the inter-particle vector ${\mathbf R}_{ij}$ into this plane (dashed line). Reproduced from Ref. \cite{Kolli15} with permission from the Royal Society of Chemistry.} 
\label{fig:hexatic_image}
\end{figure}
$\left \langle \psi_6 \right \rangle$ is equal to 1 in case of perfect hexatic ordering and otherwise vanishes.  
 The piece of information stemming from  $\langle \psi_6 \rangle$ can be supported by calculation of the average number of nearest neighbours $\langle n \rangle$ within each layer, which would be equal to 6 for perfect hexagonal order.  $\langle n \rangle$ is computed by taking for any helix the number of its nearest neighbours  in a plane and averaging this number over all helices and all configurations. Both $\psi_6$ and $\langle n \rangle$ are very sensitive to the definition of the nearest neighbour distance. We used the value corresponding to the first minimum of the radial distribution function.

We note that helices display smectic phases with or without screw-like ordering. 
We can then identify $\langle P_{1,c} \rangle$
to check whether the phase is screw-smectic or simply smectic. It turns out that the latter does not occur for all cases studied in our
simulations, while we find both screw-smectic A (Sm$_{A,S}^{*}$, no in-plane hexagonal order) and screw-smectic B (Sm$_{B,S}^{*}$, with
in-plane hexagonal order). Conversely, we do not have any evidence of a simple Sm$_A$ phase in our results for appreciable curliness ($r>0.1$).
The reason is likely
to be ascribed to the fact that before encountering a smectic phase, the system is in a screw-nematic phase, and there
is no reason for loosing this organization when initially entering into a smectic phase. The screw-organization may or may be not
lost at higher densities in the presence of short range  in-plane hexagonal order. In the former case, the helices seconday axes
$\widehat{\mathbf{w}}$ tend to be correlated within each layer but uncorrelated from one layer to the neighboring ones. We call this
organization smectic B polar Sm$_{B,p}$. In the latter case, the $\widehat{\mathbf{w}}$ are further correlated from one layer to
the neighboring ones, thus maintaining the original screw-like organization. We call this configuration screw-smectic B (Sm$_{B,S}^{*}$).

Table \ref{tab:1} summarizes the different phases exhibited by hard helices  and the relative order parameters.
\begin{table}[h]
\small
\begin{tabular*}{0.5\textwidth}{@{\extracolsep{\fill}}lll}
\hline
Phase & Code & Order parameter     \\
\hline
Nematic       & N                  & $\langle P_2 \rangle$ \\
Screw-nematic             & N$_{s}^{*}$        & $\langle P_2 \rangle$, $\langle P_{1,c}\rangle$\\
Smectic A                  & Sm$_{A}$           & $\langle \tau_1 \rangle$ \\
Screw-smectic A            & Sm$_{A,s}^{*}$     & $\langle \tau_1 \rangle$ , $\langle P_{1,c} \rangle$   \\
Smectic B                  & Sm$_{B}$           & $\langle \tau_1 \rangle$, $\langle \psi_6 \rangle $ \\
Polar smectic B            & Sm$_{B,p}$         & $\langle \tau_1 \rangle$, $\langle \psi_6 \rangle$, $\langle P_{1,c}^{\rm{layer}} \rangle$  \\   
Screw-smectic B            & Sm$_{B,s}^{*}$     & $\langle \tau_1 \rangle$, $\langle \psi_6 \rangle $, $\langle P_{1,c} \rangle$   \\
\hline 
\end{tabular*}
\caption{Summary of the different phases identified in the MC simulations of hard helices, along with the corresponding  order parameters.}
\label{tab:1}
\end{table}

\subsection{Parallel and perpendicular pair correlation functions}
\noindent In addition to global order parameters, it proves essential to introduce pair correlation functions that provide detailed insights into the structural features of the phase. In particular, it is useful to consider correlation functions resolved along the $\widehat{\bf n}$ director and perpendicular to it, according to the decoupling of the inter-particle vector shown in Eq. \eqref{eq5b}.
Thus, two relevant correlation functions are  $g_{\parallel}(R_{\parallel})$ and $g_{\perp}(R_{\perp})$, which are sensitive to inter-particle correlations along the director and perpendicular to it, respectively.

The perpendicular positional correlation function $g_{\perp}(R_{\perp})$ is related to the probability that, given a particle, another one is found 
 at a distance that, when projected onto a plane perpendicular to the director, is equal to $R_{\perp}$, and is defined as \cite{Cinacchi08}:
\begin{eqnarray}
g_\perp(R_\perp) &=& \frac{1}{2\pi R_{\perp} N} \left\langle \frac{1}{\rho  L_Z}\sum_{i=1}^{N}\sum_{j\neq i}^{N} \delta(R_\perp -| \mathbf{R_{ij}} \times \hat{\mathbf{n}}|)\right\rangle 
\label{eq6}
\end{eqnarray}
where $\delta(\ldots)$ is the Dirac $delta$ function and 
$L_z$ is the length of the simulation box along the director $\widehat{\mathbf{n}}$.
Note that the number density is taken inside the average as the  volume $V$
 of the box may change in the course of simulations.
The volume is calculated as $V=L_x L_y L_z$, where $L_x$ and $L_y$ are the lengths of the simulation box perpendicular to director $\widehat{\mathbf{n}}$.

\begin{figure}
\centering
\includegraphics[width=10cm]{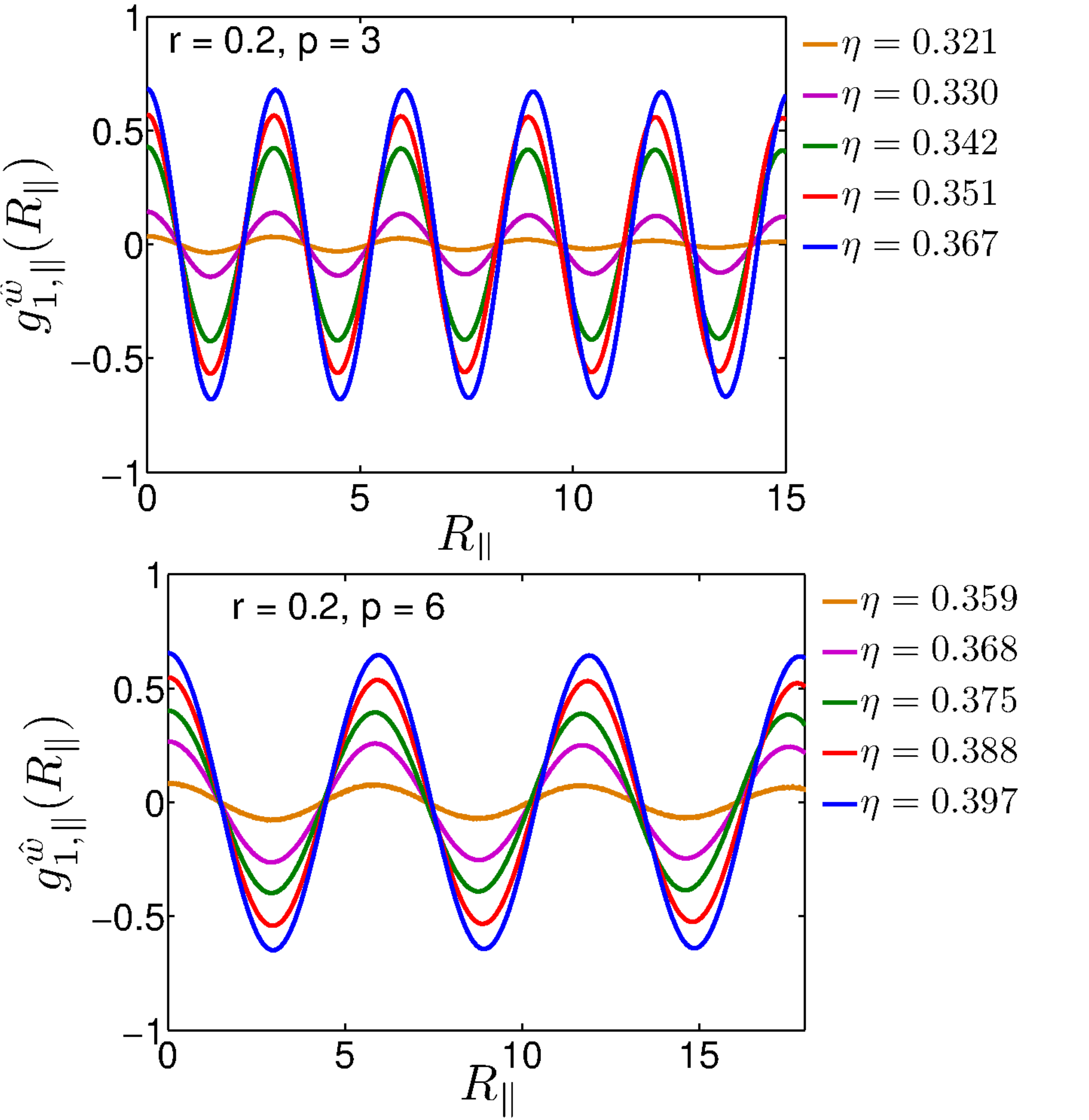}
\caption{
Correlation function $g^{\hat{w}}_{1,||}(R_{||})$ as a function of , the projection of the inter-particle distance along the director $\hat{\mathbf{n}}$, from MC simulations of hard helices with  radius $r=0.2$ and pitch $p=3$ and $p=6$, at different values of the volume fractions $\eta$.}
\label{fig:g1w_result}
\end{figure}

The parallel positional correlation function, $g_{\parallel}(R_{\parallel})$,
is related to the probability that, given a particle, another one is found on a plane perpendicular to the director $\widehat{\mathbf{n}}$, at the distance $R_{\parallel}$, and is defined as:
\begin{eqnarray}
g_\parallel(R_\parallel) &=& \frac{1}{N} \left\langle \frac{1}{\rho  L_x L_y }\sum_{i=1}^{N}\sum_{j\neq i}^{N} \delta\left(R_\parallel - \mathbf{R}_{ij} \cdot \hat{\mathbf{n}}\right)\right\rangle 
\label{eq7}
\end{eqnarray}
One important correlation function, designed to identify the onset of screw-like ordering:
\begin{eqnarray}
g_{1,\parallel}^{\widehat{\mathbf{w}}}(R_{\parallel}) &=& 
\left \langle \frac{\sum_{i=1}^N \sum_{j\neq i} ^ N \delta(R_{\parallel}-\mathbf{R}_{ij}\cdot \hat{\mathbf{n}}) (\widehat{\mathbf{w}}_i \cdot \widehat{\mathbf{w}}_j)} 
{\sum_{i=1}^N \sum_{j\neq i} ^ N \delta(R_{\|}-\mathbf{R}_{ij} \cdot \hat{\mathbf{n}})} \right  \rangle. \nonumber \\
\label{gw}
\end{eqnarray}
The function $g_{1,\parallel}^{\widehat{\mathbf{w}}}(R_{\parallel})$ detects correlations  
between the $C_2$ symmetry axes of helices with the value of its maximum coinciding with $\langle P_{1,c} \rangle$.
It also allows a direct measure of the pitch of the phase, as seen in Figure \ref{fig:g1w_result} 
for helices with $r=0.2$ and $p=3$ and $p=6$. 
The plots show  pronounced oscillations, whose amplitude increases with increasing density, with periodicity equal to the pitch $p$ of the helical particles.
The set of correlation functions help to distinguish different phases. 

In the isotropic and nematic phase, both $g_\parallel(R_\parallel) $ and $g_\perp(R_\perp)$ exhibit the characteristic behavior of 
a liquid; in the smectic A phase $g_{\perp}(R_{\perp})$ is still liquid-like, whereas $g_{\parallel}(R_{\parallel})$ 
displays the characteristic peaks of one-dimensional ordering; in the smectic B phase, $g_{\perp}(R_{\perp})$ will additionally acquires the structure characteristic
of hexatic ordering; finally, $g_{1,\parallel}^{\widehat{\mathbf{w}}}(R_{\parallel})$ 
exhibits oscillations with
periodicity identical to the particle pitch $p$ in the presence of screw-like ordering, both in the nematic and smectic phases.

\section{The physical origin of cholesteric and screw-like order}
 \label{sec:origin}
\noindent Elongated particles such as rigid spherocylinders interacting only via excluded volume form a nematic phase upon compression because of the gain in translational entropy accompanying the onset of orientational order. Rigid hard helical particles will display a similar tendency to align their main axis along a common director $\widehat{\mathbf n}$, but with two important differences.

The first difference stems from the fact that helix is a chiral object. Again,  
formation of the nematic phase is driven by gain in translational entropy accompanying the orientational order, for helices with sufficiently high aspect ratio. However, due to the particle chirality, pair configurations with 
right- and left-handed  twist of the  helix axes ($\widehat{\mathbf u}$) are not equivalent. So, they give different contributions to the average excluded volume \cite{Frezza14,Belli14,DussiJCP}. 
Depending on the morphology and the state point, right- or left-handed contributions may prevail for helices with a given handedness. 
Anyway the net effect is very small, since it derives from the unbalance of oppositely signed and very similar contributions.    
The macroscopic result is that the director is twisted; a right/left-handed cholesteric phase is formed if pair configurations with a right/left-handed twist have on average a smaller excluded volume. 
Due to the smallness of the net chirality of the average excluded volume,  the cholesteric pitch $\cal P$ is  
large on the particle length scale  (for non-chiral particles the net chirality vanishes and the N phase is formed, which can be seen as a cholesteric with infinite pitch). Moreover, a given cholesteric handedness cannot be uniquely associated with a given handedness of the constituent helices. According to Straley model \cite{Straley}
inversion of the cholesteric handedness is predicted as a function of the thread pitch $p$ for hard threaded rods:  considering right-handed screws, the cholesteric handedness would switch from right to left on moving from  tight to looser pitch. Analogous results were obtained in recent studies of 
hard helical particles, which additionally  showed that a given system may exhibit N$^\ast$ handedness inversion upon increasing density \cite{Frezza14,Belli14,DussiJCP}.

\begin{figure}
\centering
\includegraphics[width=5cm,angle=270]{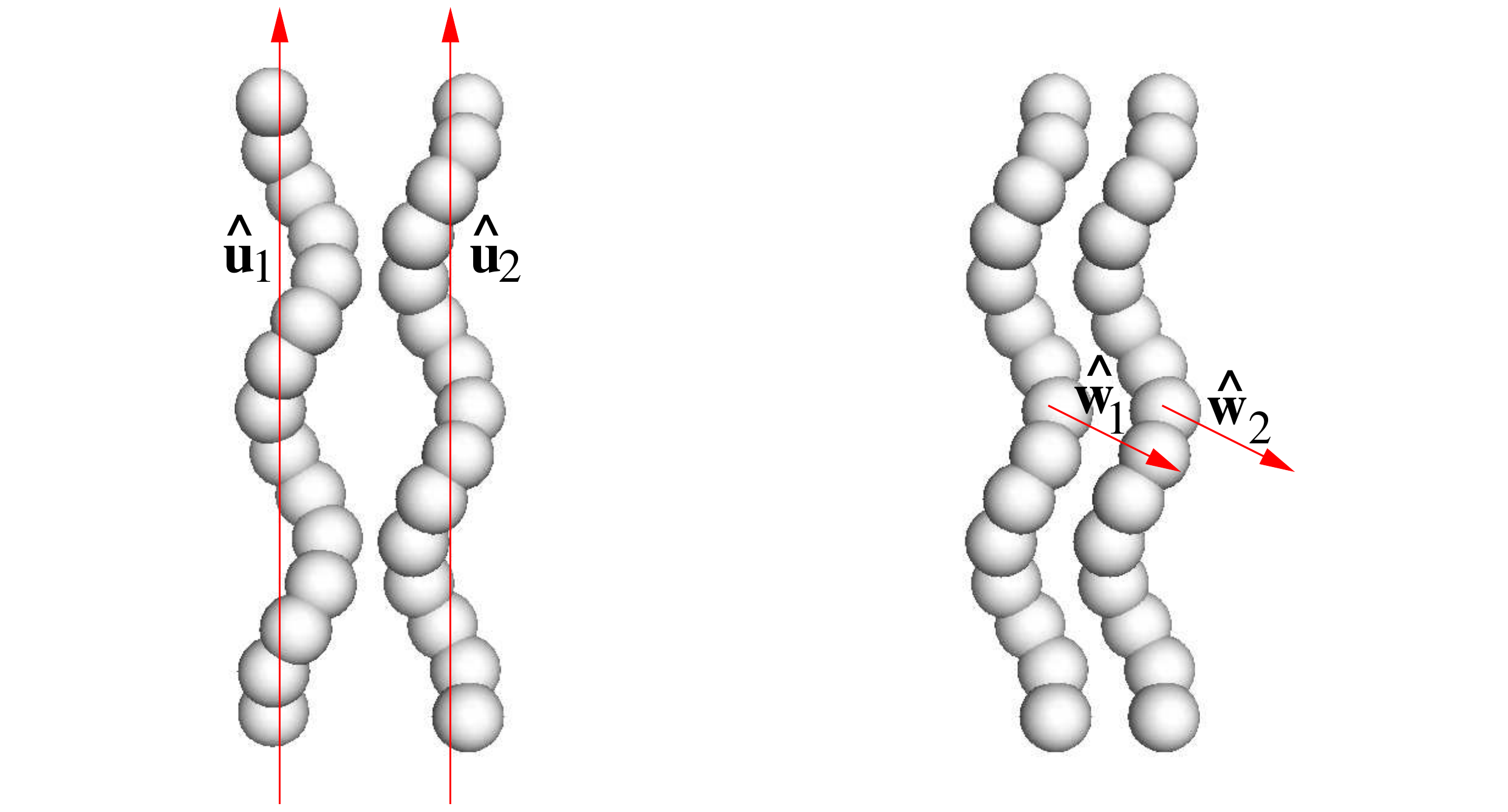}
\caption{Pairs of helices in antiphase (left) and in phase (right).}
\label{fig:helixscrew}
\end{figure}

The second important difference from rod-like particles originates from the particular shape of the helix that triggers a tendency for neighboring parallel helices in a nematic organization to slide one over another in a screw-like fashion.
This can be rationalized as follows. Imagine to have two parallel helices with no azimuthal correlation between their respective orientations in the plane perpendicular to $\widehat{\mathbf n}$ (see Figure \ref{fig:helixscrew}, left). Under these conditions, the system looks locally nematic. At higher densities, neighboring parallel helices are expected to have a significant intrusion between grooves, thus providing an effective azimuthal correlation stemming from having helices in phase (see Figure \ref{fig:helixscrew}, right), with a corresponding significant loss of rotational entropy about their own axes. This can only be compensated by a translational motion of the helix along the nematic direction $\widehat{\mathbf n}$ in a screw-like fashion. We call this phase screw-nematic phase, and denote it as N$^\ast_S$. A similar mechanism also occurs in smectic phases, that will be denoted as screw-smectic.  
Screw-like order is characterized by a phase periodicity $\cal P$ and handedness identical to the pitch $p$ and handedness of the constituent helices, at any density. Another distinctive consequence is the emergence of a local polar order ($\widehat{\mathbf c}$ is a polar director), which is missing in both N and N$^\ast$ phases.  

\section{The phase diagram of hard helices} 
\label{sec:phase}
\noindent We have performed NPT-MC simulations for helices of different morphology (see Figure \ref{fig:helixshapes}) and using the order parameters and correlation functions defined in Section \ref{sec:OP} we have identified  the sequence of phases formed by them. In the following we will present some representative results. Scaled quantities will be used throughout, with the bead diameter $D$ taken as the unit of length and $k_B T$ as the unit of energy.   
In all calculations right-handed helices were considered. For left-handed helices the same phase diagrams would be obtained, but  in the case of 
helical phases the handedness would be reversed.

A special comment has to be deserved to  cholesteric phases. For the reasons mentioned in Section \ref{sec:numerical}, no cholesteric phase could be found in our NPT--MC simulations due to the periodic boundary conditions. 
However, this does not compromise the  significance of the phase diagrams, since   the
existence of a large-scale director twist does not essentially modify the thermodynamic and ordering properties of the N phase. 
On the other hand, using an Onsager-like theory (Section  \ref{sec:DFThelical}) we calculated the cholesteric pitch in the state points corresponding to nematic phases, and we obtained $\cal P$ values of the order of 100-1000 $D$ \cite{Kolli15}.  
So, what throughout the paper, based on the  NPT--MC results, we have denoted as nematic phases, should rather be considered as actual cholesteric phases.

\subsection{The equation of state}
\begin{figure}
\centering
\includegraphics[scale=0.3]{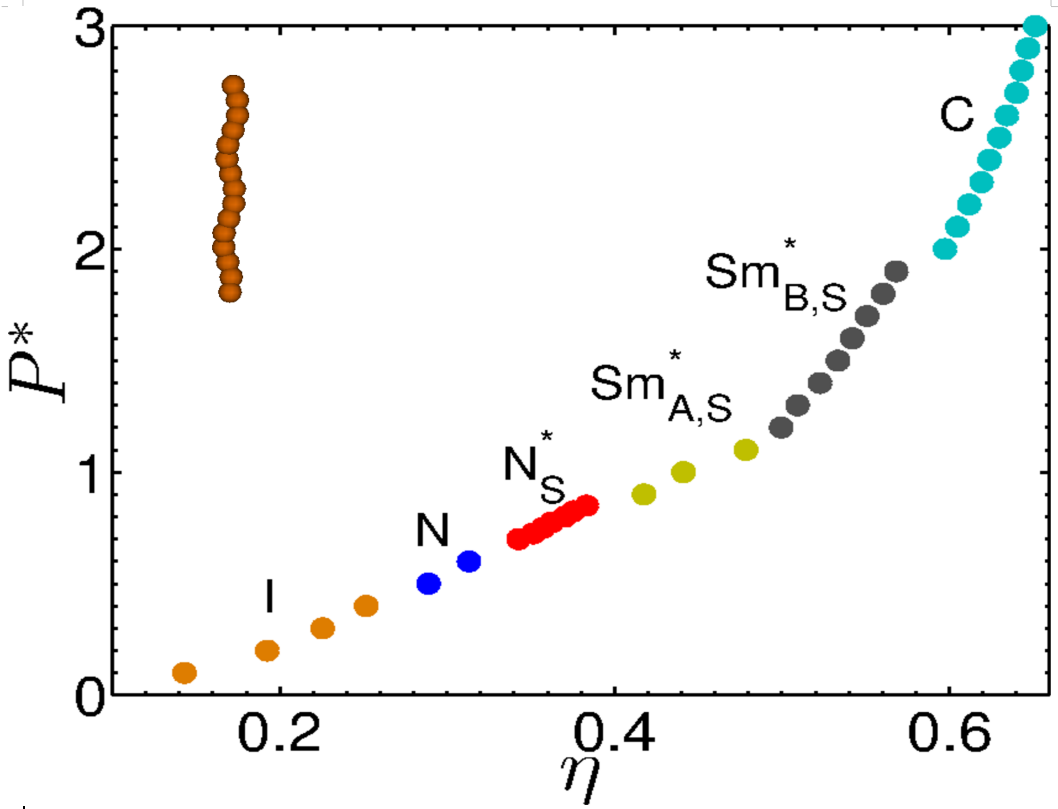}
\includegraphics[scale=0.3]{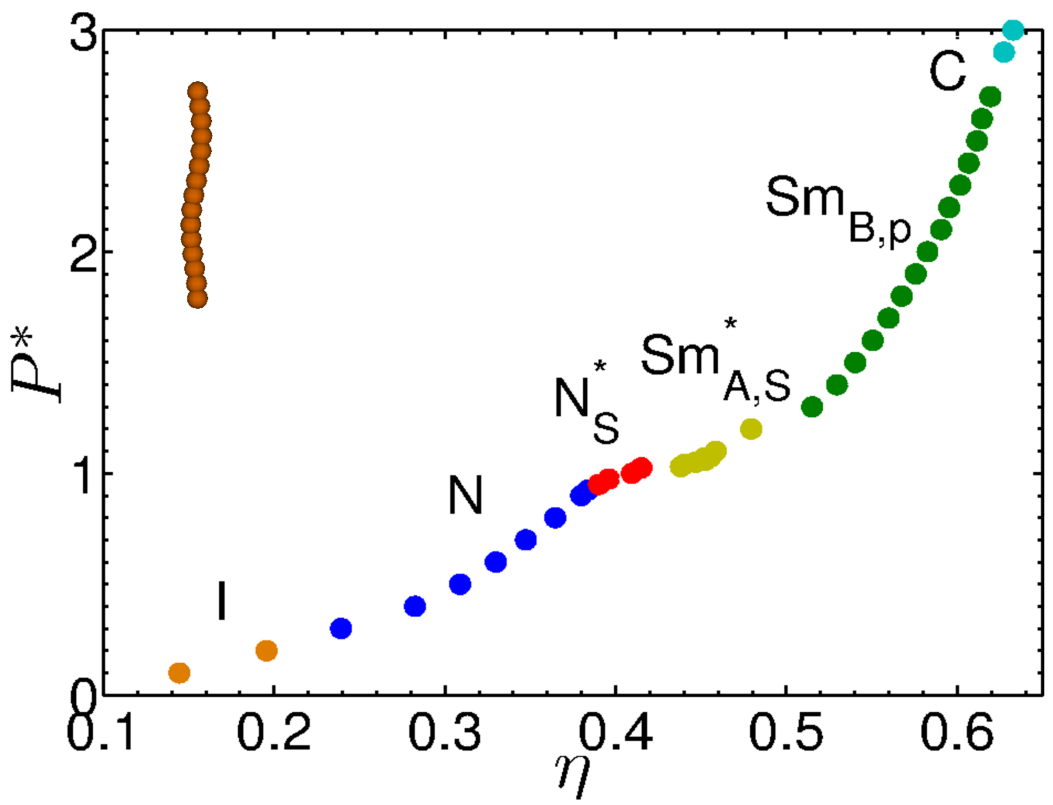}
\caption{Reduced pressure $P^\ast=P D^3 /k_BT$ versus volume fraction $\eta$ for helices with $r=0.2$ and pitch $p=4$ (top) and $p=8$ (bottom). 
The corresponding shape of the helix is displayed in the insets. Different phases are identified by different colors
and labeled accordingly.}
\label{fig:eos_r2}
\end{figure}

\noindent In the volume fraction-pressure plane the pressure displays a jump in the case of a first-order transition
or a change in slope for a continuous transition. However, the differences between liquid crystal phases can be small, and even first order transition may be very weak.  
As an example, Figure \ref{fig:eos_r2} shows the equation of state for helices with $r=0.2$ and pitches $p=4$ and $p=8$.
In both cases we can see the sequence I--N--N$^\ast_S$--Sm with increasing density. This is a common behaviour. 
The isotropic-nematic and nematic-smectic transitions are found always first-order, as indicated by the jump in the nematic  $\langle P_2 \rangle$ and smectic $\langle \tau_1 \rangle$ order parameters. 
Conversely, the transition between nematic N and screw-nematic N$_{s}^{*}$ phase appears to be continuous,
on the basis of the behavior of $\langle P_{1,c} \rangle$.
As we can see in Fig. \ref{fig:eos_r2}, the relative amplitude of the different phases depends on the helix morphology.
The N$_{S}^{\ast}$ phase occurs at higher volume fraction than the N phase, which can be rationalized considering that high 
packing is required for the screw-like mechanism to be advantageous from the entropic view point.
For  curly helices the N$^\ast_S$ phase tends to widen and  in some cases, for very curly ones, we have found direct transition from the isotropic to the screw-nematic phase.
At even higher density, the N$^\ast$ is superseded by a smectic phase which preserves the screw-like order.
In this phase, denoted as Sm$_{A,S}^{\ast}$, the centers of mass of the helices are homogeneously distributed within each layer, 
with their orientation in phase, that is the $\widehat{\mathbf{w}}$  vectors tend to be parallel, and different layers are correlated in a screw-like fashion. This is sketched in Figure \ref{fig:cartoon_smectic} (left). Upon increasing the density, we observe two alternative scenarios.

In the Sm$_{B,S}^{\ast}$ phase, Figure \ref{fig:cartoon_smectic} (center), the system 
maintains the same global organization as in the Sm$_{A,S}^{\ast}$ case,
but with additional in-plane hexatic order. 
Essentially, helices stack on top of each other 
with the appropriate azimuthal orientation to form a system of infinite helices along the
$\widehat{\mathbf{n}}$ direction, in this way having a significant gain in translational entropy.
In the Sm$_{B,p}$ phase, Figure \ref{fig:cartoon_smectic} (right), in-plane 
hexatic  and azimuthal correlations are preserved, but
different layers are fully uncorrelated. Here entropy is maximized by loosing the screw order in favour of
an optimal packing of the helices with appropriate off-sets. 

\begin{figure}
\centering
\includegraphics[width=10cm]{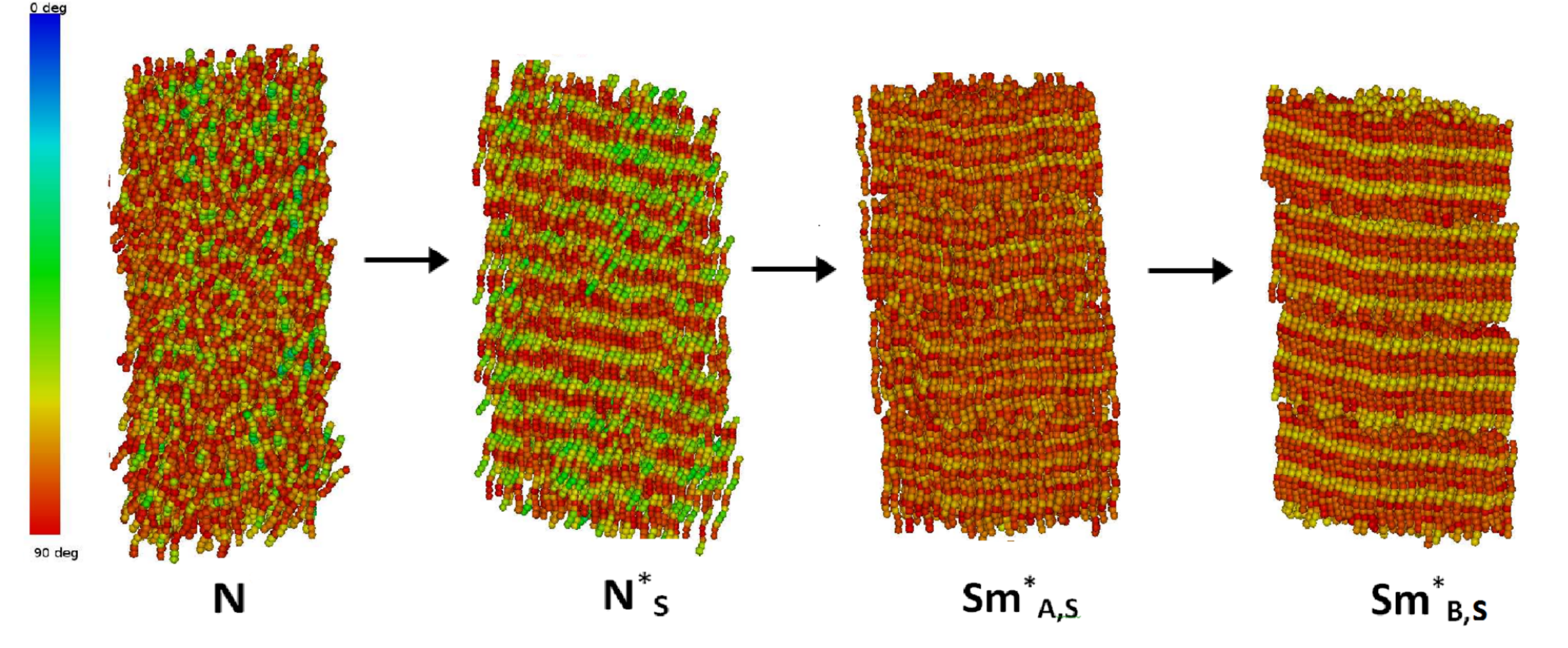}
\includegraphics[width=10cm]{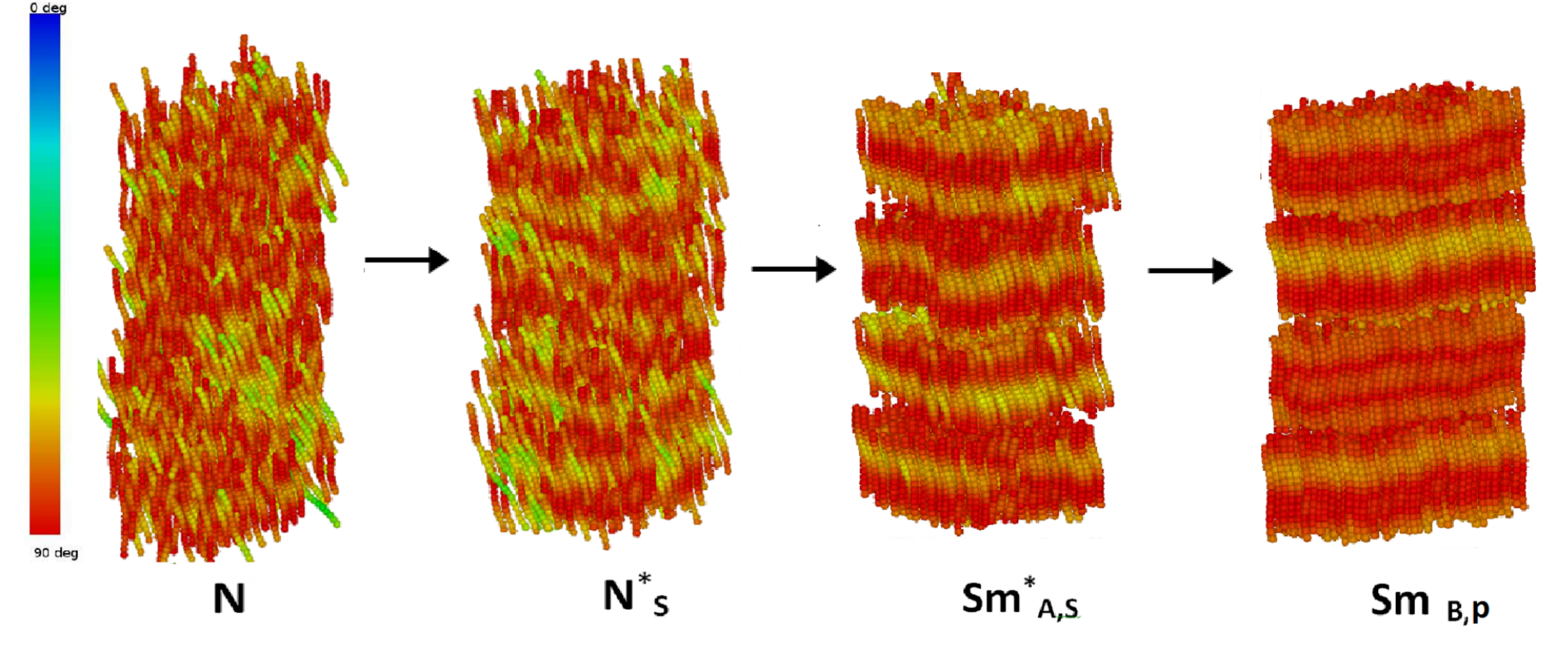}
\caption{
Snapshots from NPT--MC simulations of helices with $r=0.2$, $p=4$ (top) and $r=0.2$, $p=8$ (bottom)
Density increases on moving from left to right displaying the sequence of phases. Color is coded according to the projection of the local tangent to helices onto a plane perpendicular 
to the director $\mathbf{\widehat n}$ \cite{qmga}. 
Adapted from Ref. \cite{Kolli15} with permission from the Royal Society of Chemistry.}
\label{fig:snapshots}
\end{figure}

\begin{figure}
\centering
\includegraphics[width=3cm]{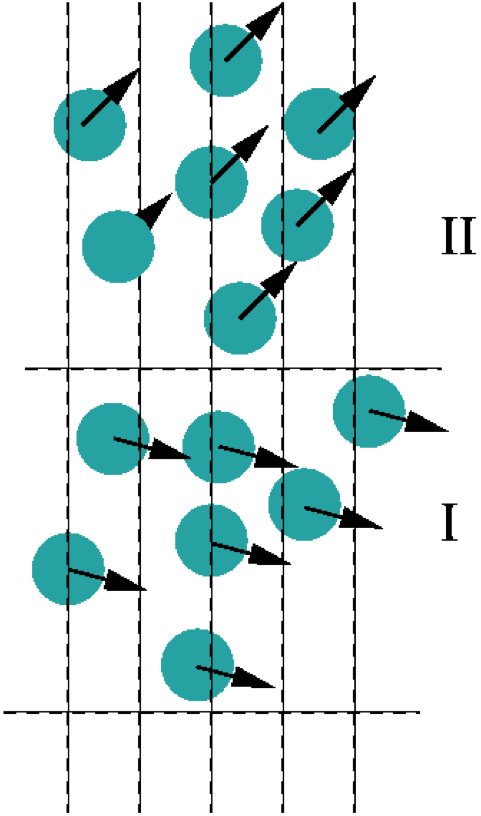}
\includegraphics[width=3cm]{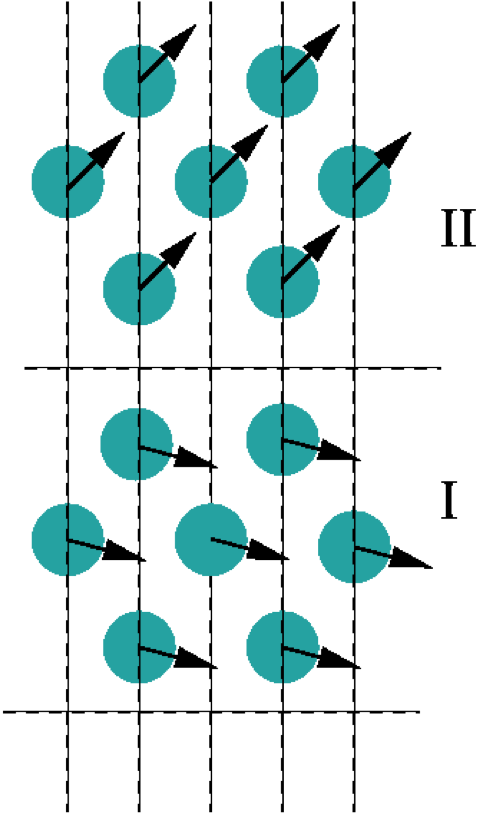}
\includegraphics[width=3cm]{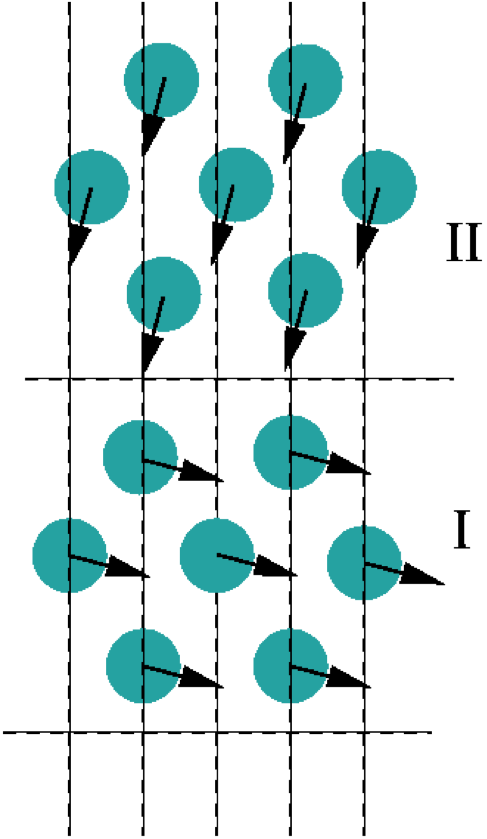}
\caption{Cartoon of a top view of two neighboring smectic layers (I) and (II) with different in-plane 
organizations. Sm$_{A,S}^{*}$ (left), Sm$_{B,S}^{*}$ (center) and Sm$_{B,p}$ (right).}
\label{fig:cartoon_smectic}
\end{figure}

The different kinds of smectic phases are shown in Figure \ref{fig:snapshots}.

\subsection{Phase diagrams in the volume fraction-pitch plane}
\label{subsection:rpscrew}

\noindent We are now in the condition to summarize the phase behaviour of different helical shapes. This will be accomplished in the volume fraction-pitch plane for the two different radii discussed in previous sections ($r=0.2$ and $r=0.4$). The case $r=0.1$ will be also 
shown as a paradigmatic example of a very slender shape not showing any tendency to a screw-like organization.

\subsubsection{Phase diagram for $r = 0.1$} 
\begin{figure}
\centering
\includegraphics[width=8cm]{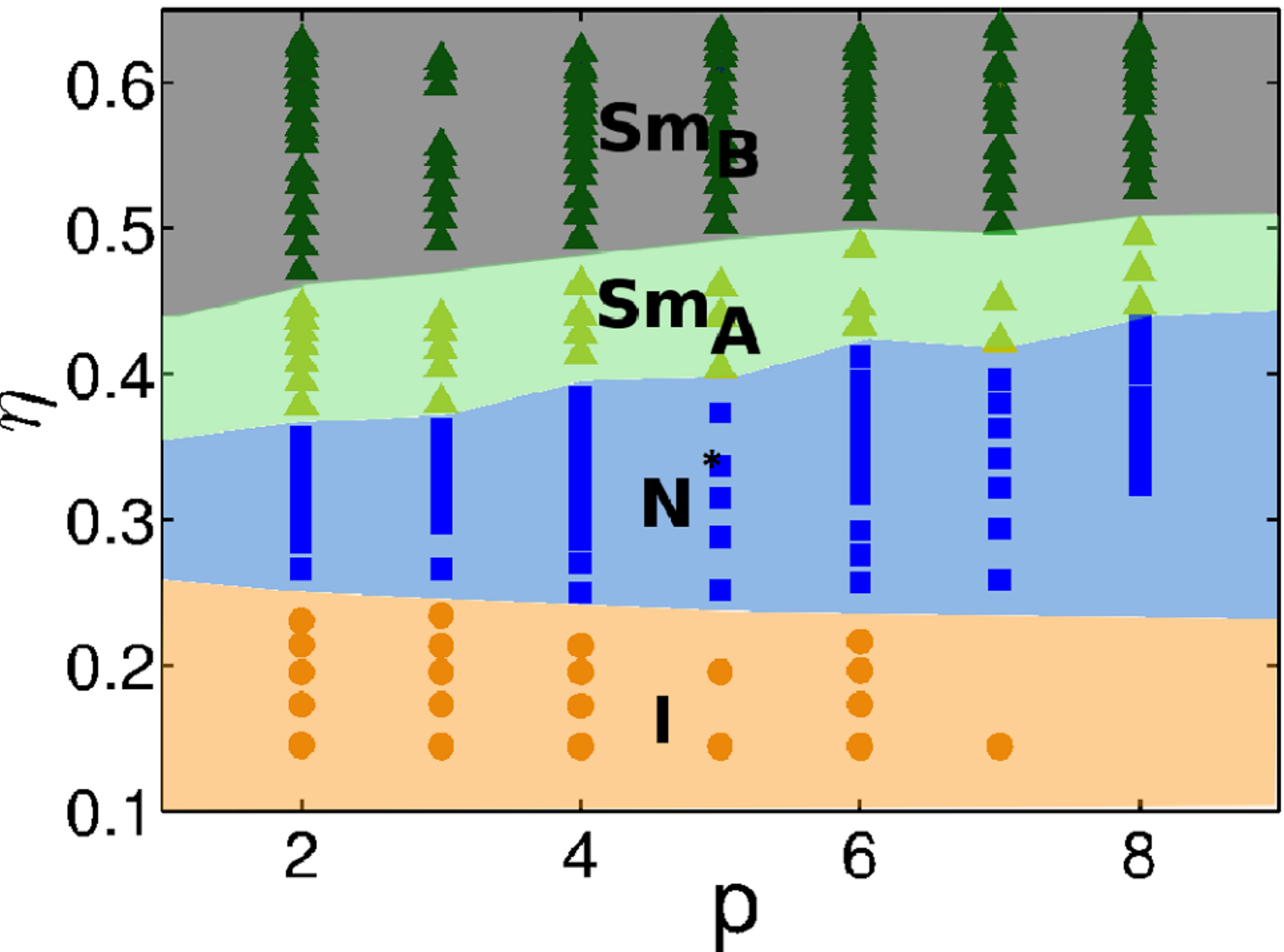}
\caption{Phase diagram in the plane volume fraction $\eta$ \textit{versus} pitch $p$ for helices having $r = 0.1$. 
Symbols correspond to calculated points. Different phases are labeled as indicated in Table I. The symbol N$^\star$ is used for state points that in NPT-MC simulations were found in the nematic phase, and then their cholesteric pitch was determined by an Onsager-like theory. Reproduced from Ref. \cite{Kolli15} with permission from the Royal Society of Chemistry.}
\label{fig:r1_shades}
\end{figure}

\noindent As visible in Fig. \ref{fig:helixshapes},
helices with $r = 0.1$ are almost rod-like particles with  
high aspect ratio. Therefore we expect a 
phase behaviour not very different from that of their spherocylinder counterparts. Figure \ref{fig:r1_shades} shows this to be the
case: as in spherocylinders, we find isotropic to nematic N and nematic to 
smectic A and B transitions upon increasing $\eta$, the difference between Sm$_A$ and Sm$_B$ being the 
hexatic order of the latter, as discussed. 

\begin{figure}
\centering
\includegraphics[width=8cm]{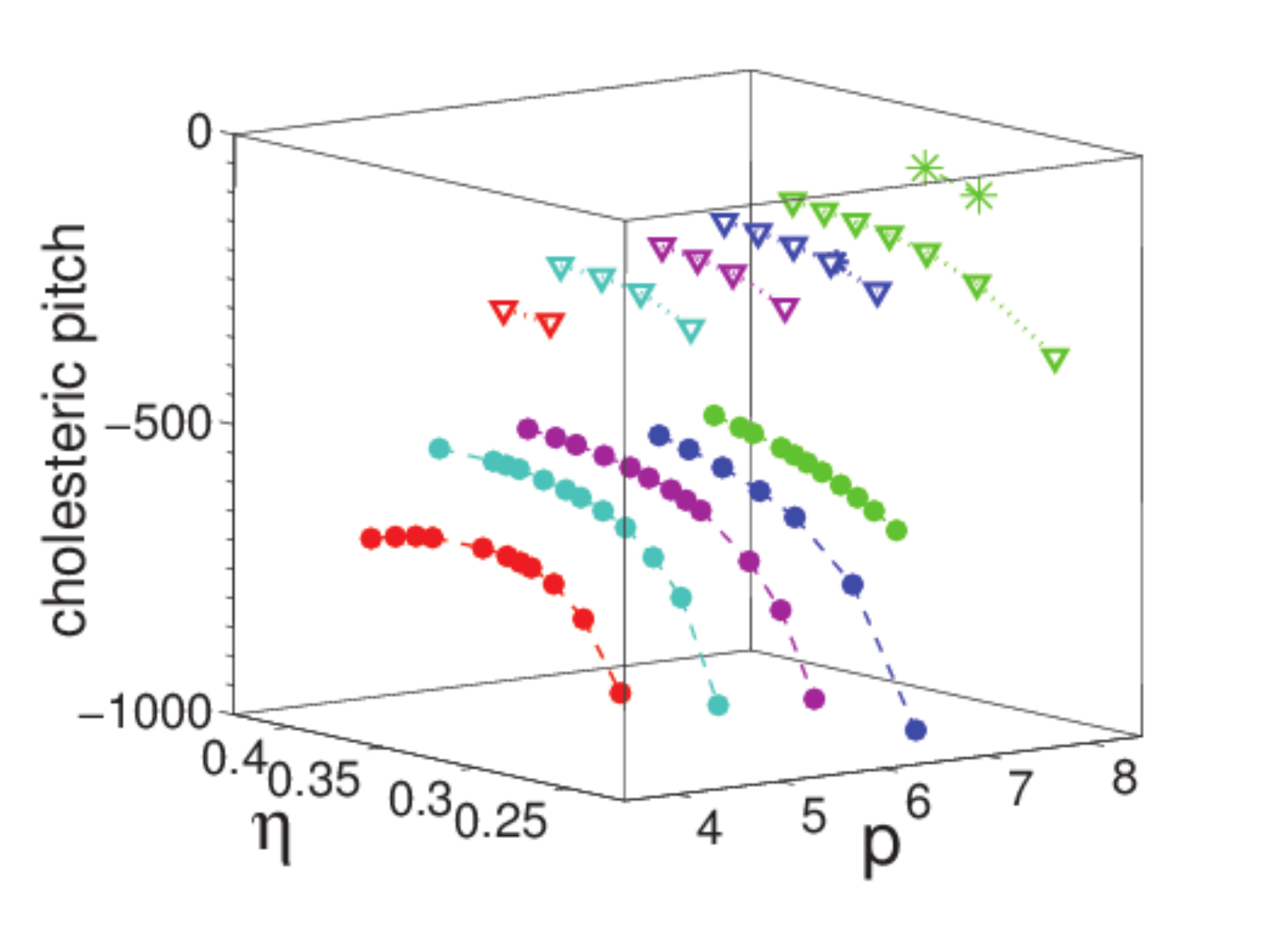}
\caption{Cholesteric pitch $\cal P$, calculated in state points indicated as N$^{\ast}$ in the phase diagrams in Fig. \ref{fig:r1_shades} ($r=0.1$, full circles), Fig. \ref{fig:r2_shades} ($r=0.2$, open triangles) and Fig. \ref{fig:r4_shades} ($r=0.4$, asterics). 
$\eta$ is the volume fraction and $p$ is the pitch of the helical particles (only $p$ values greater than 3 are shown).}
\label{fig:a}
\end{figure}

The only qualitative difference from  the phase diagram of hard spherocylinders is 
in the  nematic phase, which  here is cholesteric rather than uniform nematic, as
shown by  the values of $\cal P$ predicted by an Onsager-like theory and shown in Figure \ref{fig:a}. 
Left-handed N$^\ast$ phases are predicted (${\cal P} < 0$),  
with pitches 
longer than 500 $D$, which tend to decrease with increasing density.

The absence of screw-like order 
in this case is clearly due to the weak curliness of the helices, which  is not sufficient to trigger the screw-like behavior. 
The aspect ratios are not significantly different from each other, ranging between $8.78$ and $9.14$. As the pitch $p$ increase, 
the aspect ratio slightly increases and stabilizes the nematic phase at lower $\eta$, consistently with the spherocylinder
counterpart. 
\subsubsection{Phase diagram for $r = 0.2$} 
\begin{figure}
\centering
\includegraphics[width=8cm]{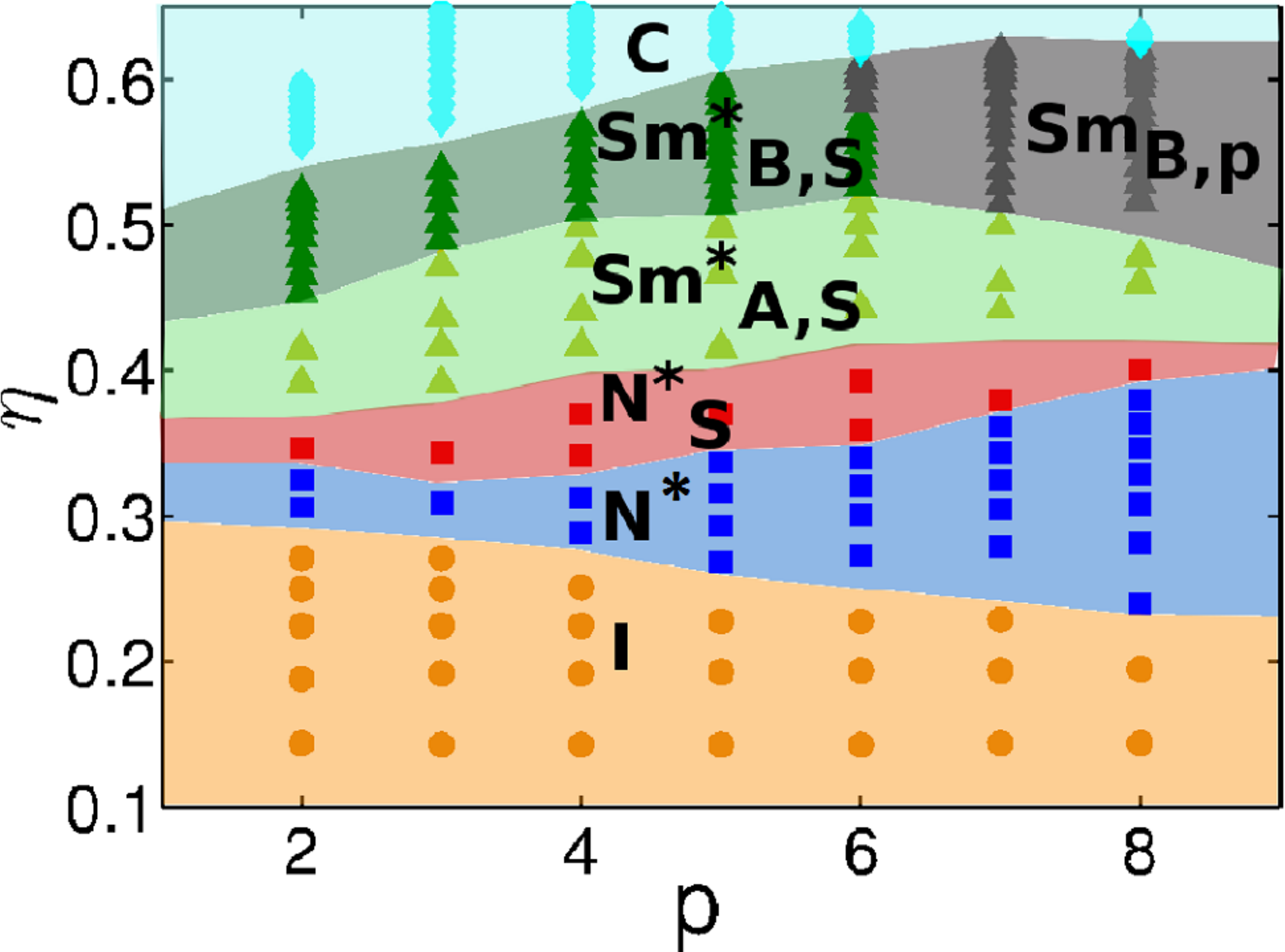}
\caption{Phase diagram in the plane volume fraction $\eta$ \textit{versus} pitch $p$ for helices having $r = 0.2$. 
Symbols correspond to calculated points. Different phases are labeled as indicated in Table I; C is a high density compact phase. The symbol N$^\star$ is used for state points that in NPT-MC simulations were found in the nematic phase, and then their cholesteric pitch was determined by an Onsager-like theory. Reproduced from Ref. \cite{Kolli15} with permission from the Royal Society of Chemistry.}
\label{fig:r2_shades}
\end{figure}

 \noindent Figure \ref{fig:r2_shades} shows the phase diagram of helices with $r=0.2$.
Comparing with the case $r=0.1$ we can appreciate significant differences, with the presence of several new phases, most of which chiral.
As in the case $r=0.1$, at a given density these systems exhibit a transition from the isotropic to the nematic phase. And again, calculations based on an Onsager-like theory predict this to be a twisted nematic phase. The pitch, shown in Figure \ref{fig:r2_shades}, is much smaller than for $r=0.1$, sign of a stronger chirality of interactions between the curlier helices. Then at higher density the screw-nematic phase appears, which is followed by smectic phases specific of helices. The increase of helix radius  from $r=0.1$ to $r=0.2$ brings about a significant curliness in helical shape to promote a screw-like ordering at higher densities. 

An important difference occurs for $p$ values below and above $p=6$. 
For $p<6$ the Sm$_{A,s}^{\ast}$ develops into a Sm$_{B,s}^{\ast}$ ordering upon increasing $\eta$. Here, the screw-like coupling between different layer is then favoured as 
hexatic ordering is gradually setting in. Conversely, for $p>6$ this is lost in favour of a rearrangement of neighboring layers
to achieve an optimal packing. A glance back to Figure \ref{fig:helixshapes} reveals the reason for this being
again related to the fact that for $p>6$ helices are so slender to make unfavourable the screw-like organization.
As we will see below, this will not be the case for $r=0.4$.  
\subsubsection{Phase diagram for $r = 0.4$} 
\begin{figure}
\centering
\includegraphics[width=8cm]{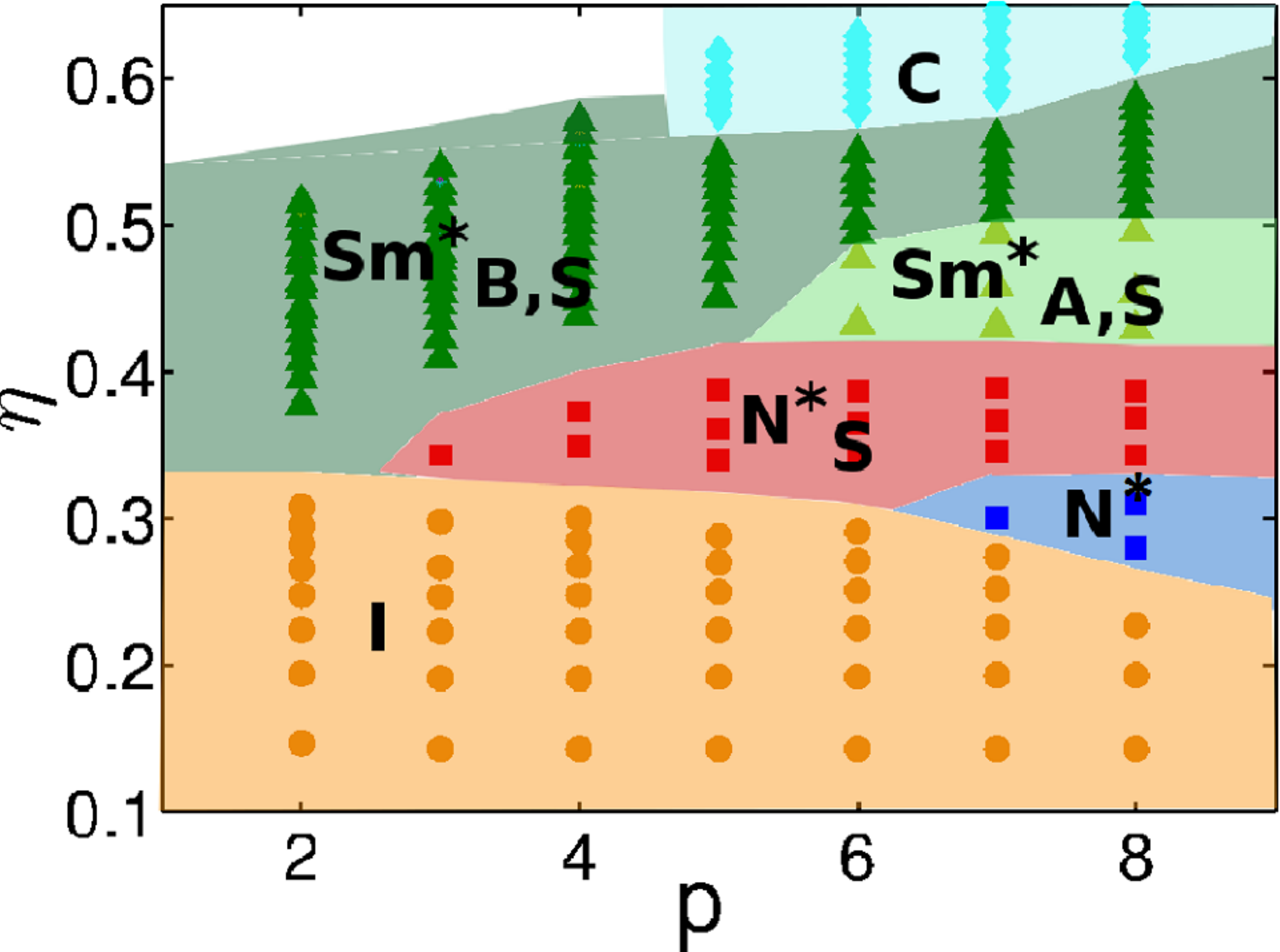}
\caption{Phase diagram in the plane volume fraction $\eta$ \textit{versus} pitch $p$ for helices having $r = 0.4$. 
Symbols correspond to calculated points. Different phases are labeled as indicated in Table I; C is a high density compact phase. The symbol N$^\star$ is used for state points that in NPT-MC simulations were found in the nematic phase, and then their cholesteric pitch was determined by an Onsager-like theory. Reproduced from Ref. \cite{Kolli15} with permission from the Royal Society of Chemistry.}
\label{fig:r4_shades}
\end{figure}

\noindent Figure \ref{fig:r4_shades} shows the phase diagram of helices with $r = 0.4$. 
A notable difference with respect to the case $r=0.2$ hinges on the increased stability of the N$_{S}^{\ast}$ phase with respect to the N$^{\ast}$ counterpart, and the disappearance of 
the Sm$_{B,p}$ phase in favour of a wider Sm$_{B,S}^{\ast}$ phase that
extends over all $p$ values above a certain $\eta$. 
This was anticipated above, and can be ascribed to the curliness of all helices with $r = 0.4$, 
as can be inferred again from  Figure \ref{fig:helixshapes}.
Another difference with the $r=0.2$ case is given by the presence of several triple points, a signature of a rich
polymorphism.
At higher densities, one finds rather compact structures, denoted as ``C'', with the exception of 
helices with low $p$,  for which the situation is less clear. Here one could expect the presence of rotator phases in analogy with what
occurs in the spherocylinder counterpart. 
The N$^{\ast}$ phase appears only in a narrow region, for helices with a long pitch. According to an Onsager-like theory, in this phase the director is spontaneously twisted, with a pitch $\cal P$ smaller than in the systems previously examined.

\section{Helical (bio)polymers and colloidal particles }
\label{sec:exp}
\noindent
It is interesting to see how results obtained for hard helices compare with the behavior of systems made of helical particles, several examples of which can be found in natural or synthetic polymers 
and colloids.
Cholesteric phases have been found in solutions of stiff or
semi-flexible, covalent or supramolecular, helical polymers, either in
water or in organic solvent.
In such systems  a variety of interactions is generally present 
and this, in addition to the intrinsic flexibility of polymers, 
 may  affect the features of the cholesteric phase \cite{Katsonis}, as well as the relative
stability of phases \cite{Cinacchi08}. The presence of linear
self-assembly, which introduces a state dependent length dispersion of
polymers, makes the situation even less straightforward in the case of
supramolecular polymers.
The clear picture of the phase behavior that we have reached 
represents none the less a useful, general reference, previously lacking, even for  
systems which may be much more complex. Indeed, the complexity of real
systems calls for clear guidelines, as can be obtained from  simple models
which can be fully understood.

Aqueous solutions of double stranded B-DNA ($> 100$ base pairs), which is
right-handed, form a left-handed cholesteric phase with a pitch $\cal P$ of the
order of some micrometers \cite{Rill,Livolant}.
Relatively short DNA duplexes ($< 20$ base pairs), whose aspect ratio
would be too low to induce liquid crystal order, also exhibit N$^\ast$
phases upon self-assembling into long helical aggregates, promoted by
end-to-end stacking interactions  \cite{Bellini}. In this case, however,
both right- and left-handed  cholesteric phases were found, depending on
the oligonucleotide sequence and length. Based on Straley model
\cite{Straley}, a right-handed N$^\ast$ phase would be predicted for the
B-DNA helix morphology \cite{Proni}, a prediction confirmed by
calculations based on Onsager theory for a coarse grained B-DNA model
\cite{Tombolato05, FrezzaSM}. This is also in agreement with the results
obtained for our model hard helices, which  indicate the formation of a
right-handed N$^\ast$ phase in the case of right-handed helices with small
pitch $p$ \cite{Frezza14,Belli14,DussiJCP}. 
The variety of behaviors exhibited by B-DNA oligomers, with  prevalence of left-handed cholesterics, were
ascribed to the competing effect of other interactions, specifically electrostatic
ones, superimposed to steric repulsions, which would 
promote opposite twist of particles \cite{FrezzaSM, Wensink14}.

Other helical polymers that in organic solvents form cholesteric phases
are polypetides, and particularly well studied is the case of
poly($\gamma$-benzyl-L-glutamate) (PBLG). This has a right-handed
$\alpha$-helical structure and may form either a right- or a left-handed
N$^\ast$ phase, depending on the solvent, with pitches of the order of
hundreds of nanometers.
In this uncharged system a special role of dispersion interactions was
invoked, whose effect would depend on the relative dielectric constant of
polymer and solvent \cite{Samulski}, and again on the competition  
between dispersion interactions 
the underlying steric repulsions \cite{Osipov, Emelyanenko03}.

Cholesteric phases have been observed also in helical colloidal systems: a
well known example is that of filamentous viruses, which are formed by a
DNA core wrapped by a coating of helically arranged proteins.  \textit{fd}
and M13 viruses were found to form  left-handed N$^\ast$ 
phase with pitch  $\cal P$
ranging from tens to hundreds of micrometers \cite{Grelet}; however
the \textit{fd} Y21M mutant, which differs from   \textit{fd} only for
having a methionine in place of a tyrosine as the 21rst aminoacid in the
coat protein, forms a right-handed N$^\ast$ phase with a pitch almost an
order of magnitude larger  \cite{Barry09}. Either the competition between
steric and electrostatic interactions \cite{Grelet06} or a key role of
helical shape fluctuations \cite{Grelet} were proposed for these systems,
but no comprehensive understanding has been reached, yet.

Experimental evidence of screw-nematic order has been observed
in concentrated suspensions of helical flagella isolated from
\textit{Salmonella typhimurium} \cite{Barry06}.
While flagella with a rod-like shape exhibit a nematic phase,
filaments with a pronounced helical character were found to undergo a
direct transition from the isotropic to
a  modulated nematic phase with pitch $\cal P$ in the micrometer scale, which
in the original paper was denoted as conical.
It can be easily verified that the experimental results are fully
compatible with what we have called screw-nematic phase, with the helix
axes ($\widehat{\mathbf{u}}$) preferentially aligned along the same
direction
(the main director $\widehat{\mathbf{n}}$) and the two-fold symmetry axes
of helices $\widehat{\mathbf{w}}$ spiralling around
$\widehat{\mathbf{n}}$. This was discussed in detail  in Refs.
\cite{Kolli14a,Kolli14b, Kolli15}.
Direct transition from the isotropic to the N$^\ast_s$ phase can be seen
in the phase diagram that we have calculated for helices having $r=0.4$
and various values of the pitch (Figure \ref{fig:r4_shades}). In such a phase diagram the
N$^\ast_s$ phase is superseded by smectic phases at higher density, at
variance to the experimental system, which  did not exhibit any smectic
phase.
This difference can be ascribed to the  length polydispersity of the
helical flagella, which inhibits the formation of layers, but is fully
compatible with the existence of screw-like order.

It is also worth  noting that in the experiments on
helical flagella, no cholesteric phase has been observed.
It would be interesting to investigate in detail what is the role
of length polydispersity in the stability of a cholesteric phase 
in systems of helical particles.

To our knowledge, there has been no evidence of screw-like order in
polymeric systems. A possible reason is that  polymers, owing to their
shape and flexibility, do not meet the requirements for the stabilization
of phases with screw organization. This is what occurs for the slightly
helical particles whose phase diagram is shown in Figure \ref{fig:r1_shades}. Another
possible reason is the difficulty in experimental detection of space
modulations with pitch $\cal P$ in the nanometer range. We hope that new,
targeted experiments can shed light on this issue, possibly involving the design and  synthesis of 
polymers with a more pronounced helical shape.


\section{Conclusions and perspectives }
\label{sec:conclusions}
\noindent In this Chapter we have tried to account for some recent results obtained by
considering a fluid of rigid helices interacting only sterically, that is via excluded volume only.

To this aim, we have used a combination of Monte Carlo simulations and density functional theory.

Notwithstanding the fundamental role of helical motifs in nature, a detailed study of this fluid was surprisingly missing, likely because
phase behaviour of helical particles was assumed to be a minor variation of that of hard straight rods. Conversely, we found a rich and unconventional 
polymorphism of entropically driven liquid crystal phases that, besides to conventional nematic and smectic (A and B) phases, also appearing
in a fluid of rod-like particles,  display  specific phases, whose origin we have discussed.

A rigid helical particle is a chiral object, in such  left-handed helices cannot superimposed to their right-handed counterparts. Because of
this chirality, neighboring parallel helices in a nematic phase experience a twist that propagates in the direction perpendicular to the
original plane following a helical path with a pitch typically much larger than the particle itself. Therefore the phase looks nematic on
a short scale but is cholesteric on a much larger scale. We found this phase to be favoured at moderate volume fractions and moderate
curliness.

At larger densities and for more pronounced curliness, when there is sufficient interlocking between neighboring helices to promote azimuthal coupling, 
a different mechanism sets in, where there is a tendency for the parallel helices to slide up and down along the nematic director to increase their
translational entropy compensating the corresponding loss of rotational entropy due to the azimuthal coupling. This is the driving
mechanism for the formation of a screw nematic phase whose origin is rooted in the helical shape of the particles.

We have further discussed how a similar mechanism favours the formation of screw-smectic phases (both A and B), in addition to a another smectic phase
(called `` B polar'') where this organization is replaced by a different one with horizontal sliding and rotation of neighboring smectic layers which differs from the conventional  Smectic B 
for the presence of polar transversal order. 

These findings suggest a significant sensitivity of the liquid crystal phases to the shape of the helices that is to be accounted for
in all those experimental systems for which hard helices can be reckoned as a good minimal model.
This includes biological systems, such as helical flagella, but also  helically nanostructured materials exhibiting special optical properties that are
of interest for photonic metamaterials.

The above model could be made more realistic by adding charges, localized interactions, by allowing mixing right- and left-handedness, as well as by including
flexibility in the helices. These model variants will be useful in describing different systems.

Yet, there are a number of possible further studies even 
maintaining the helices as hard and rigid.
One interesting issue hinges on the possible presence of columnar phases in a system of monosized hard helices, 
which are exhibited for instance by \textit{fd}-viruses \cite{Grelet15}.
Equally deserving a dedicated study is a detailed analysis of the high-densities ordered phases, as they might provide some surprising features
in view of the non-convex nature of the helices \cite{Damasceno15}. 
Other important, still unexplored, issues concern the behavior of mixtures of helices, differing in length and morphology, mixtures of enantiomers, as well as
the dynamics of the onset of screw-like phases.
  
All these issues will be pursued in the near future.

\begin{acknowledgments}
G.C. thanks the Government of Spain for
the award of a Ram\'{o}n y Cajal research fellowship and
the financial support under the grant FIS2013-47350-C5-1-R and the grant MDM-2014-0377. A.F. and A.G. acknowledge financial support from
MIUR PRIN-COFIN2010-2011 (contract 2010LKE4CC).
The use of the SCSCF multiprocessor cluster at the Universit\`{a} Ca' Foscari Venezia
is gratefully acknowledged.
\end{acknowledgments}

%
\bibliographystyle{apsrev}

\end{document}